\documentclass[preprint,3p,onecolumn,pra,aps,showpacs,superscriptaddress,article]{elsarticle}
\usepackage[english]{babel}
\usepackage[utf8]{inputenc}
\usepackage{amsmath,amssymb,amsthm}
\usepackage{xcolor}
\usepackage{float}
\usepackage[export]{adjustbox} 
\usepackage{textcomp}
\usepackage{amssymb}
\usepackage{graphicx}
\usepackage{esint}
\usepackage{lineno,hyperref}
\usepackage[toc,page]{appendix}
\usepackage{tabularx}
\def\<{\left\langle}
\def\>{\right\rangle}
\def\({\left(}
\def\){\right)}
\def\dis{\textrm{dis}}
 \journal{Physica A}
\DeclareUnicodeCharacter{2217}{*} 
\begin{document}

\begin{frontmatter}

\title{Segregation in spatially structured cities}

\author{Diego Ortega\corref{cor1}}
\ead{dortega144@alumno.uned.es}
\address{Dto. Física Fundamental, Universidad Nacional de Educación
a Distancia (UNED), Spain}

\author{Javier Rodríguez-Laguna}
\address{Dto. Física Fundamental, Universidad Nacional de Educación
a Distancia (UNED), Spain}

\author{Elka Korutcheva}
\address{Dto. Física Fundamental, Universidad Nacional de Educación
  a Distancia (UNED), Spain}
\address{G. Nadjakov Institute of Solid State Physics, Bulgarian
  Academy of Sciences, 1784 Sofia, Bulgaria.}

\begin{abstract}
Half of the world population resides in cities and urban segregation is becoming a global issue. One of the best known attempts to understand it is the Schelling model, which considers two types of agents that relocate whenever a transfer rule depending on the neighbor distribution is verified. The main aim of the present study is to broaden our understanding of segregated neighborhoods in the city, i.e. \textit{ghettos}, extending the Schelling model to consider economic aspects and their spatial distribution. To this end we have considered a monetary gap between the two social groups and five types of urban structures, defined by the house pricing city map. The results show that ghetto sizes tend to follow a power law distribution in all the considered cases. For each city framework the interplay between economical aspects and the geometrical features determine the location where ghettos reach their maximum size. The system first steps shape greatly the city's final appearance. Moreover, the segregated population ratios depends largely on the monetary gap and not on the city type, implying that ghettos are able to adapt to different urban frameworks.

\end{abstract}

\end{frontmatter}


\section{Introduction}
The world's population is expected to increase by 2 billions in the next 30 years  \cite{UN2019}. Therefore, if urban segregation is now a major issue, its importance will be aggravated in the near future.  

Ghettos constitute the main expression of segregation. A ghetto is a part of a city in which members of a minority group live, especially as a result of political, social, legal, environmental, or economic pressure \cite{ghetto}. Regrettably, this issue affects millions of people across the world. These urban dwellers live in areas whose facilities have lower standards than the rest of the city. The common feature of these zones is the impoverishment of the area. Even though segregation can have different origins, we focus our attention on the racial one, which is strongly linked with an economic inequality \cite{britannica}. 

One of the first theoretical works in the segregation field was put forward by T.C. Schelling \cite{Schelling}. He proposed a model comprised of two social groups (\textit{red} and \textit{blue}) which occupy a certain lattice leaving some vacancies. As people tend to seek out neighbors who are similar to themselves, large clusters of the same type of agents are typically created. Schelling introduced a \textit{tolerance} parameter, i.e. the number of neighbors of a different type that an agent can withstand in her/his neighborhood remaining \textit{happy}.  If the agent is unhappy a relocation into a vacancy spot takes place whenever a certain \textit{transfer rule} is verified. 

Despite its simplicity, this model and its extensions have attracted great interest. We can classify its variants into two types, depending on the main source of motivation: physics or socioeconomic research. In the former category we have the thermodynamics approach, such as the evaluation of the specific heat or susceptibility \cite{closed}, a connection with the Blume-Emery-Griffiths (BEG) model \cite{BEG} was established in \cite{open}, its statistical properties analyzed  \cite{Dallasta08}, the vacancy border between clusters classified as belonging to the Edward-Wilkinson class \cite{ortega_r} and its roughening and diffusion mechanisms characterized \cite{Albano12}. Even a physical analogue of the Schelling model was developed \cite{vinkovic_2006}. Broadening our perspective, works whose main motivation stemmed from economics or social science usually include economic terms associated with a housing market \cite{zhang2004,portugali_95,portugali_2000}. Some works have been able to predict the relocation of higher-status households in suburban zones \cite{fossett_2006}. In recent years focus has shifted to complex agent based modeling (ABM) including ethical issues: works where agents are able to consider the neighborhood evolution \cite{houy19} or the influence of altruistic agents which greatly affects the final state of the system \cite{jensen18}. The balance between cooperative and individualistic dynamics was analyzed in \cite{grauwin_2009} and the inclusion and effect of fair agents is discussed in \cite{flaig_2019}. In our previous work \cite{ortega_g}, the happiness of the agents depended on the kind of neighbors and the monetary level, including and economic gap between agents of different types. This manuscript also considered an \textit{open city} approximation where agents can leave or enter the system. However, the main aim of this paper was to model the avalanche phenomena produced over the economically deprived type of agent. All housing prices were equal and the economic gap term was increased progressively until there were no blue agents left. 

In this article we address one of the most relevant open problems in the field of urban dynamics: the characterization of ghettos. We consider their formation, their size distribution and how they are affected by economic inequality and city structure.

Our work takes into account the housing market which allows us to define the city structure. In addition, the budget difference between the segregated group (blue agents) and the rest of urban dwellers (red ones), is considered in the economic gap term. Therefore, in our interpretation, clusters of blue agents represent ghettos. Agents present a \textit{happiness} level which depends on their location inside the city, the economic gap and the combination between the tolerance level and neighbors distribution. Relocation is possible if it increases or maintains their happiness. Given that we adopt an open city framework, agents can also move outside the city.

We consider cities that can be grouped into three categories: \emph{flat} cities, a model with no house pricing gradient, \emph{vertical} cities for which the house pricing depends on their distance to a coast or a river, like Barcelona \cite{bissacco_2019} or Los Angeles \cite{angeles_2015}, and \emph{radial} ones, where the local house prices depend on their distance to the center. We study three particular cases of the radial model: the first one shows an expensive suburban area, like Chicago during the 1930's, the second one exhibits an expensive city center and the last one adds internal barriers to the expensive city center model. These last two models can be associated to actual cities such as Paris or London. 

Ghetto sizes, positions and evolution are characterized for the five city types. Ghetto sizes tend to follow a power law distribution. When there is no monetary gap between the agent types, larger clusters are created near the center of the lattice, independently of the considered city type. As this gap increases, their position is displaced towards more affordable house pricing zones, while their sizes shrink due to finite-size effects. Evolution of the population is also analyzed. Clusters are created during the first time-steps, if we start from a random initial configuration. Since these clusters tend to be very stable the system final state depends largely on the events during this initial transient. We have found that final segregated population ratios are strongly influenced by the economical factors,  but, interestingly, they present a weak dependence on the city structure.

The paper is organized as follows. In Section \ref{sec:model} we  define the agent satisfaction level and link this expression with the Blume-Emery-Griffiths model, also explaining the system dynamics and detailing the five city types considered. In Section \ref{sec:results} we consider three different aspects of the population dynamics: size of segregated clusters (\ref{subsec:SCS}), location (\ref{subsec:PSC}) and evolution  (\ref{subsec:CE}), comparing the results for the different city models. Finally, our main conclusions and proposals for further work can be found in section \ref{sec:conclusion}.


\section{Model}

\label{sec:model}
Let us consider two different social groups (\emph {red} and \emph{blue}) living in a city defined by an $N\times N$ square lattice. The ratio of different neighbors in the vicinity of any agent is given by the \emph{diversity fraction}, $f_d$:  

\begin{equation}
  f_d=\dfrac{N^d}{N^s+N^d},
  \label{eq:1}
\end{equation}
where $N^s$ and $N^d$ are, respectively, the number of the same (s) and different (d) agents in the neighborhood. We consider their eight closest neighbors (\emph{Moore} neighborhood). This number is reduced for agents on the system edges, since the lattice is defined with open boundary conditions which take into account the finite size of cities. A key parameter of the system is the tolerance $T$,  the maximum fraction of different agents in their neighborhood that an agent can tolerate while remaining happy. This value acts as a threshold. In the Schelling model an agent is happy if $f_d \leq T$.

\begin{figure}[H]
\centering
\begin{tabular}{ccc}
\includegraphics[width=4.0cm, frame]{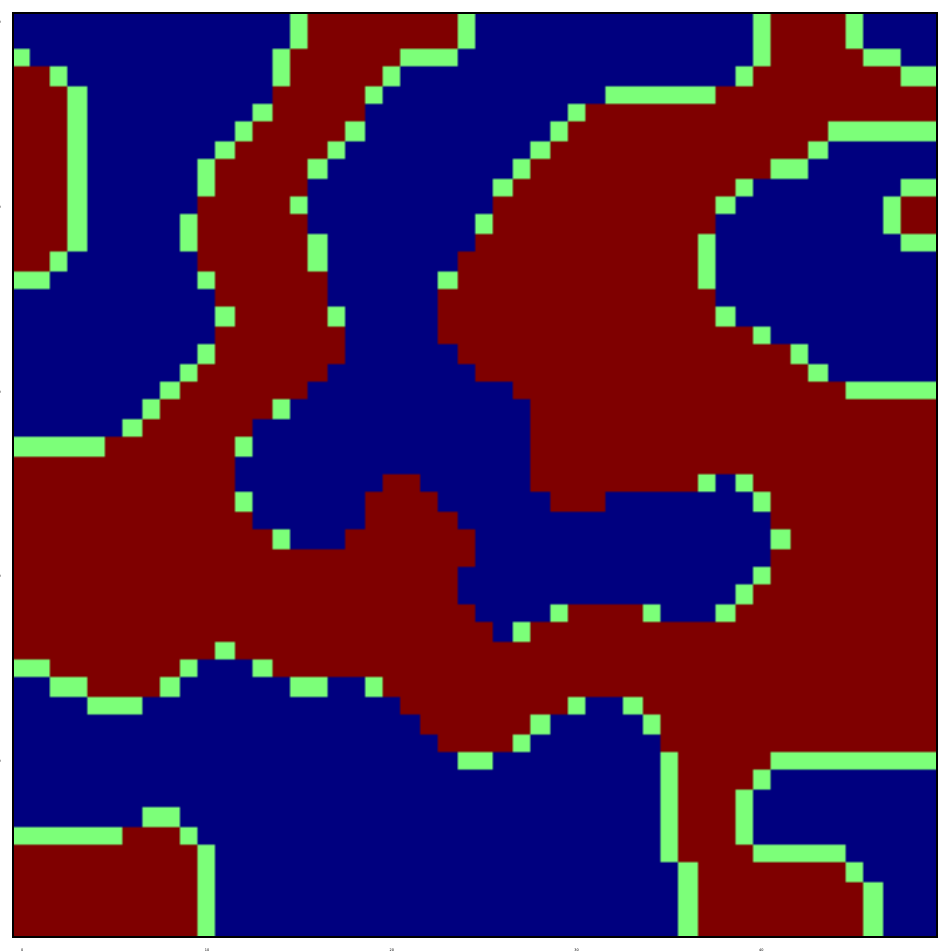} &
\includegraphics[width=4.0cm, frame]{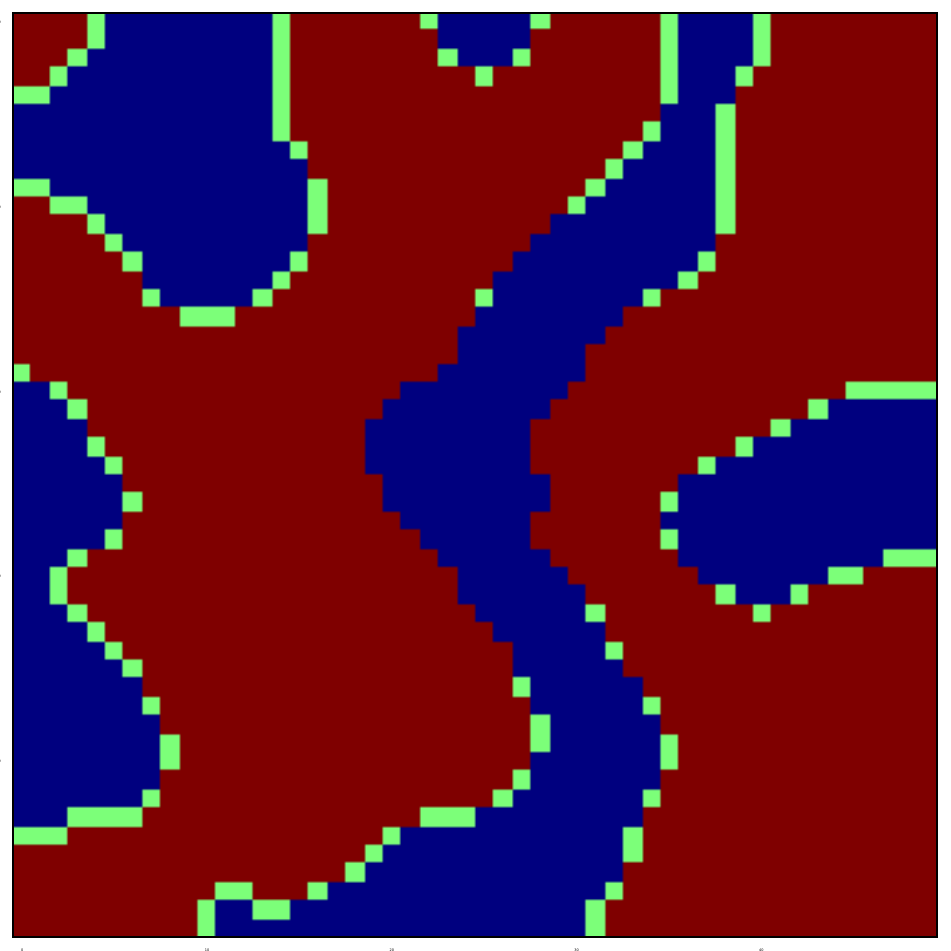} &
\includegraphics[width=4.0cm, frame]{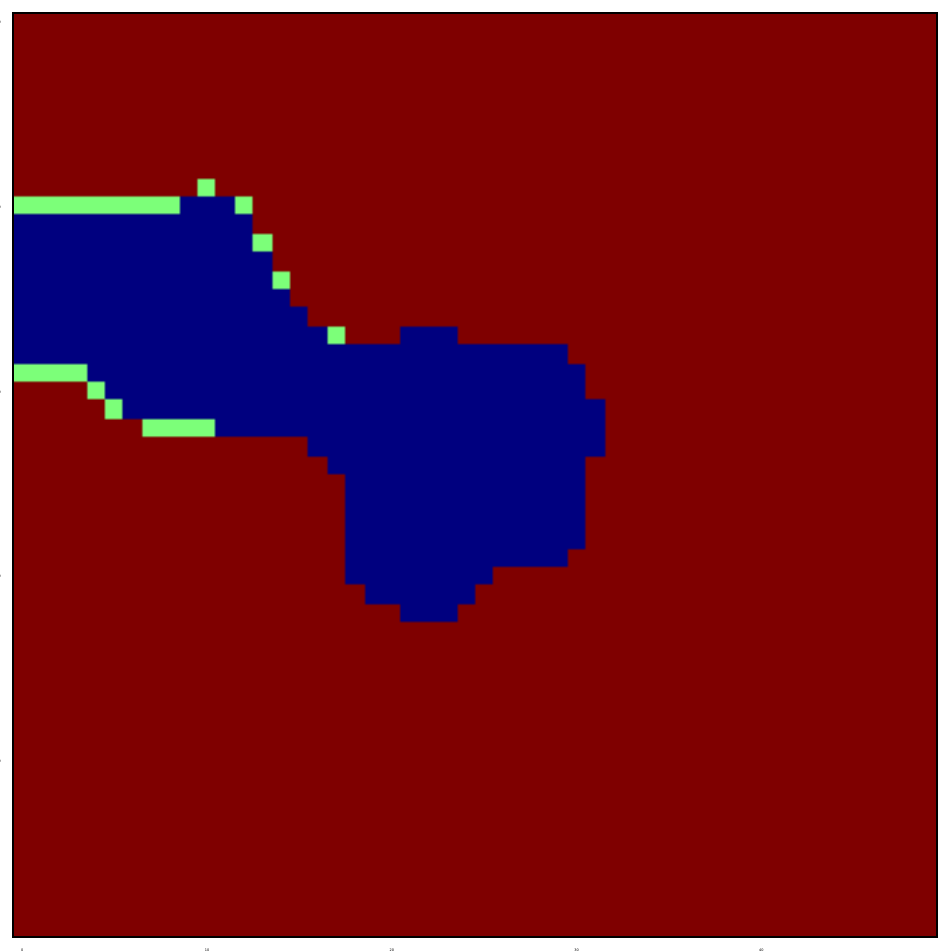} \\
(a) & (b) & (c)\\
\end{tabular}
\caption{Snapshots of the system final state for $H=0, 0.125$ and $0.250$ in (a), (b) and (c), respectively in the suburban model, see Sec. \ref{subsec:CT}. The tolerance value equal for all the agents, $T=0.25$. Red and blue agents are represented by red and blue cells, respectively. Green squares are vacancies. Blue clusters represent ghettos.}
\label{fig1}
\end{figure}

However, in order to link our physical model to social realities we measure the lack of happiness of the agent in the cell \emph{i} via the \emph{dissatisfaction index}, $I^{\dis} _i$:

\begin{equation}
  I^{\dis}_i=N^d_i-T[N^s_i+N^d_i]+ D_i + H_i,
  \label{eq:2}
\end{equation}
here, the first two terms are related to social preferences while  $D_i$ is the economic level associated to the lattice cell where the agent is located, i.e housing price. Meanwhile, $H_i$ can be understood as half the economic gap between both social groups, $H(i)=\pm H$ for blue and red types, respectively.  Obviously, when $H$ increases, the economical advantage of red agents, $-H$, over the economically handicapped blue ones, $+H$, also rises. Hence the presence of blue agents in the lattice decreases when $H$ grows, as we can see in Fig. \ref{fig1}.  It should be pointed out that lower values of $I^{\dis}_i$ are associated with happier agents. Therefore $D_i<0$ implies an affordable housing price. Contrarily, when $D_i>0$, the cost of living has risen up and the system can be understood as hostile or expensive.  Since lower $I^{\dis}_i$ values implies happier agents, the condition for satisfaction can be defined as:

\begin{equation}
   I^{\dis}_i\leq 0.
  \label{eq:3}
\end{equation}

As we can see in Eq. \eqref{eq:2} the total dissatisfaction index also depends on economic variables, measured by the housing price and the membership to a particular social group. For example, agents which are surrounded by neighbors of the opposite kind may remain if the effective economic level is friendly enough. 

When the tolerance $T$ is constant our framework can be mapped into a Blume-Emery-Griffith model \cite{BEG}, that was introduced to study the behaviour of He$^3$-He$^4$ mixtures. We consider a spin-1 variable on each site. In our interpretation of the BEG model, spin values $s_i=-1,+1,0$ will be associated with {\em blue} agents, {\em red} agents  and vacancies, respectively. The Hamiltonian \textbf{H} can be written as:

\begin{equation}
  \textbf{H}=-\sum_{\<i,j\>}\(\mathcal{J}\,s_{i}s_{j}
  +\mathcal{K}\,s_{i}^{2}s_{j}^{2}\)
  +\sum_i \(\mathcal{D}_{i}\,s_{i}^{2}+\mathcal{H}\,s_{i}\),
\label{eq:4}
\end{equation}

where $\<i,j\>$ stands for the eight nearest neighbors. This Hamiltonian represents a spin-1 model with coupling constant $\mathcal{J}$, biquadratic exchange constant $\mathcal{K}$, local crystal field $\mathcal{D}_{i}$ and an external magnetic field of intensity $\mathcal{H}$. The associated values for our segregation model in this framework are $\mathcal{J}=1$ and $\mathcal{K}=2T-1$. Local crystal and magnetic fields correspond to the $\mathcal{D}_i=2D_{i}$ and $\mathcal{H}=2H$ parameters. Details on these calculations can be found in \cite{ortega_g}. Comparing Eq. \eqref{eq:2} and Eq. \eqref{eq:4} we can observe that while the housing price $D_{i}$ maps into a chemical potential related to the position of the $i$ cell, the economic gap $H$ role is played by the magnetic external field and related to the spin. In other words, if $D_{i}>0$ the system reduces its energy by expelling agents, and if $H>0$ the system reduces its energy either expelling blue agents or attracting red ones. In this way the minimizing principle that guides the evolution of the Hamiltonian in Eq. \eqref{eq:4} is equivalent to the increasing of the happiness level from Eq. \eqref{eq:2} that includes social preferences and economic terms.

\subsection{System dynamics}

We start from a random initial configuration with equal proportions of vacancies, red and blue agents, i.e. $1/3$ of the lattice cells for each kind. We study an open city model in which agents can enter or leave the city depending on their dissatisfaction level. 

At each iteration we choose an internal or external interchange with equal probabilities. On one hand, if the selected change is internal, two spots $i$ and $j$ are randomly chosen. Nevertheless, $i$ must be occupied by an agent, while $j$ must be a vacancy. If the interchange increases or maintains the agent happiness, i.e. $I^{\dis}_i\geq I^{\dis}_j$ as can be calculated from Eq. \eqref{eq:2}, the move is accepted and the agent in the $i$ place is relocated into $j$. On the other hand, if the interchange is external, a random spot $i$ is selected. Now, two outcomes are possible depending on whether the cell is empty or occupied. If it is an empty site we try to occupy it with an agent of a random type, considering equally likely the two social groups. The move is carried out if the selected agent becomes satisfied in the spot, i.e. $I^{\dis}_i\leq0$, as we saw in Eq. \eqref{eq:3}. If the selected spot $i$ is occupied, the agent leaves the system if $I^{\dis}_i>0$.

$N\times N$ iterations of this algorithm give rise to a single time-step, as it is customary in Monte-Carlo approaches.
\subsection{City types}

\label{subsec:CT}

From now on, we use two notations for the house pricing: the previous one, $D_i$ referred to \emph{i} cell, and $D(i_x,i_y)$, related to the lattice coordinates. We switch between them depending on the context.

\begin{figure}[H]
\centering
\begin{tabular}{cccc}
\includegraphics[width=3.7cm, height=3.0cm,frame]{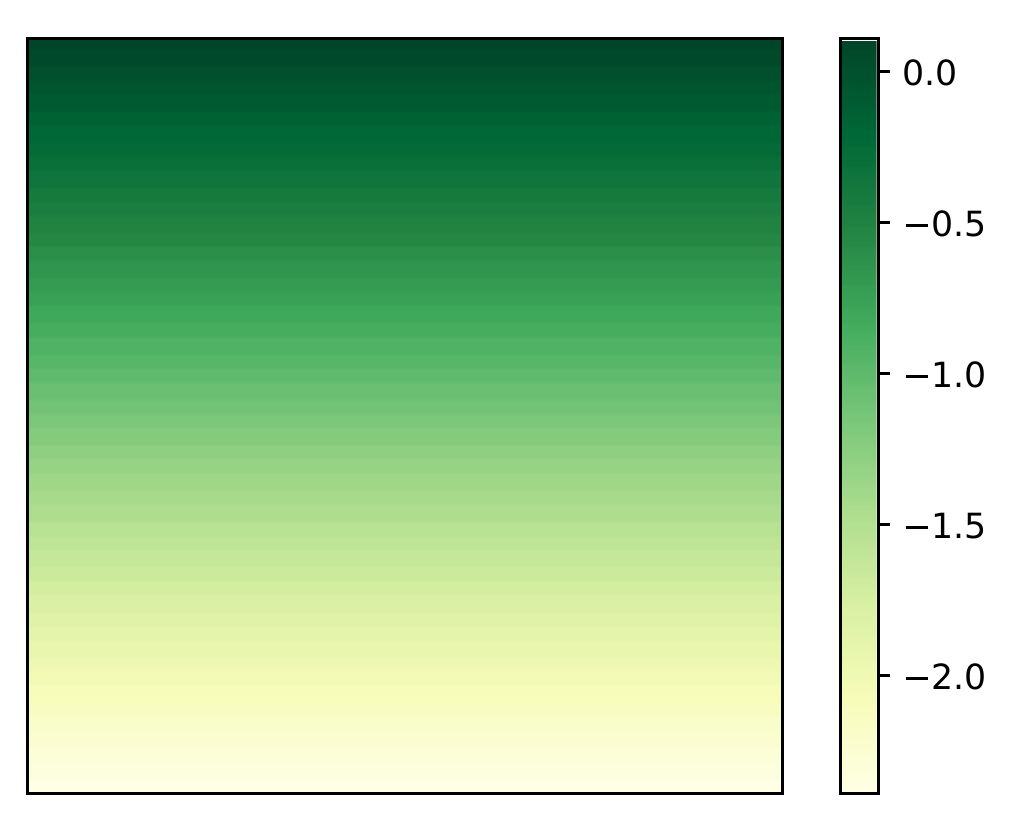}&
\includegraphics[width=3.7cm, height=3.0cm, frame]{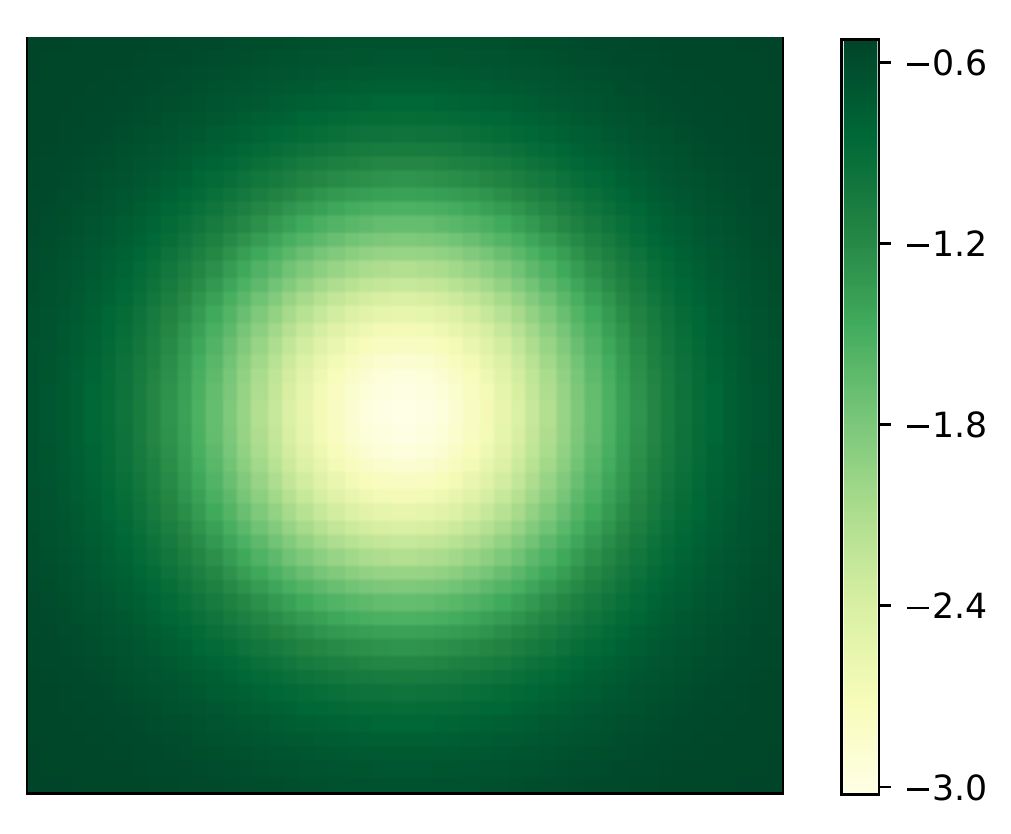}&
\includegraphics[width=3.7cm, height=3.0cm,frame]{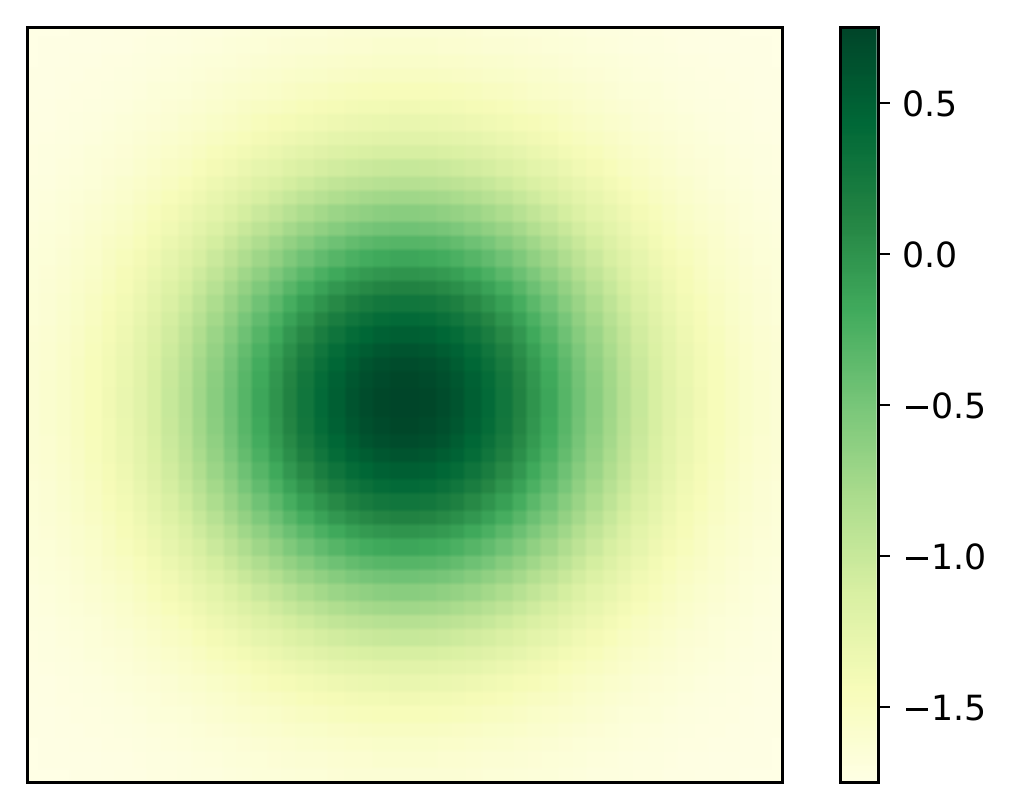}&
\includegraphics[width=3.7cm, height=3.0cm, frame]{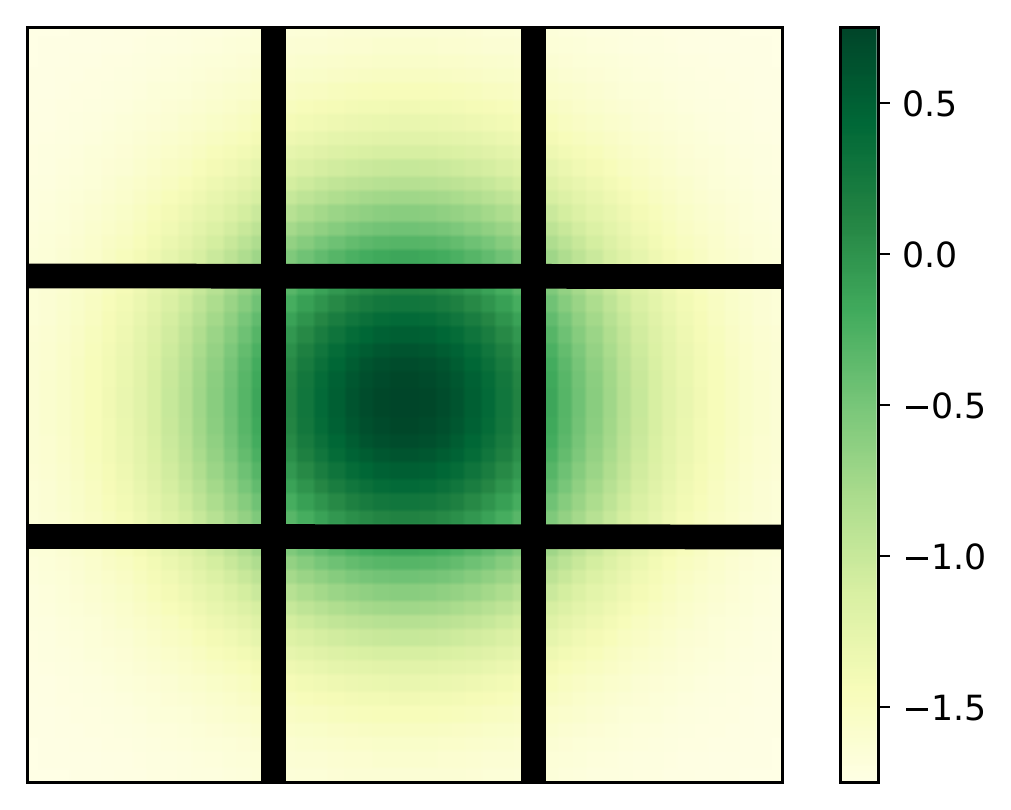}\\
a)&b)&c)&d)\\
\end{tabular}
\caption{Distributions of the house pricing $D(i_x,i_y)$ in vertical a) suburban b) core c) and grid d) cities. Brighter tones are associated with low-priced areas contrarily to darker colours which involve more expensive zones.}
\label{D}
\end{figure}

We model our city structures through the spatial variation of the house pricing, which is reflected in the range for $D_i$ which describes the different economic realities that take place in the urban environment. On one hand for $H=0$ and $T=0.25$, values of $D_{i} > 0.75$ imply that we find ourselves in the \emph{predominant vacancy state}: an area so expensive that it is not possible for the community to settle in \cite{ortega_g}. On the other hand, when $D_i < -2.00$, the area is so economically interesting that there are no vacancies. As the city must be attractive for the people who reside in it we choose a range between both values as it is illustrated in Fig. \ref{D}

It is also interesting to define house pricing distributions with similar averages and deviations in order to compare the differences between them. Average house prices can be calculated as $\overline{D_i}=\sum_{i_x,i_y}D(i_x,i_y)/(N \times N )$ where $i_x,i_y$ take values from 1 to N.  We have chosen the parameter values of the housing price distribution to obtain $\overline{D_i}\approx-1.14$ and standard deviations in a close range, from $0.64$ to $0.74$, following a normalization process. Therefore, all our models are comparable and we shift our focus towards geometrical aspects. 

The simplest city framework defined in our work is the homogeneous one. This \textit{flat} city shows the same house pricing in all their places, $D_{i}=-1.14$. The rest of the models can be grouped into two frameworks: radial (circular) and vertical (coastal) city types. While radial cities show an outward gradient in the house pricing from the center to the border, in vertical ones the prices vary from top to bottom.

Vertical cities are usually coastal towns which have expanded inland from the shore, thus creating a gradient in the house pricing as it can be seen in Fig. \ref{D} a). This distribution can be observed in Barcelona \cite{bissacco_2019} or Los Angeles \cite{angeles_2015}. In many cases, the coastal area tends to be the most expensive (top). As we move away from the beach, real state tends to be more economical (bottom). The coordinate along the coast is defined as $i_x$, while $i_y$ is associated to the distance to the shoreline. Thus, we define house pricing $D(i_x,i_y)$ as:
\begin{equation}
  D(i_x,i_y)=\dfrac{A_{v}}{N-1}(1-i_y)+B_v,
  \label{eq:5}
\end{equation}
where $i_x,i_y,j=1,2...N$, $A_v=2.5$ and $B_v=0.11$. The selected values for $A_v$ and $B_v$ give rise to $D(i_x,i_y)$ values lying in the range $[-2.39,0.11]$. As it was previously explained, these values define a variety of house pricing zones ranging from economically interesting areas where there are no vacancies left to expensive locations with empty spots \cite{ortega_g}. House pricing remains constant in the horizontal direction $i_x$ as we can deduce from Eq. \ref{eq:5}. 

Radial city planning dates back to many centuries ago.  Early cultures used the circular configuration to bring together the community while keeping out invaders and other dangers from the environment \cite{schwartz_radial}.  We explore three possibilities for house pricing markets: the first one is characterized by an expensive suburban zone, the second one shows a high-priced city center and the last one adds internal borders to the costly center model as it can be observed in Fig. \ref{D}b), Fig. \ref{D}c) and  Fig. \ref{D}d), respectively.  From here on we will refer to them as \textit{suburban}, \textit{core} and \textit{grid} city types following the previous order.

The suburban framework is based on the concentric rings model \cite{burgess}. In this representation, despite the existence of a small and expensive region of the city center, several concentric zones were progressively occupied by a transition zone with commercial buildings, another one with the working class apartments, and, as we move away from the center, better and more expensive residential areas to be inhabited. The applicability of this model to actual cities is a complex issue. It was applied to describe the Chicago urban area in the 1930 decade, but it was considered deprecated until recently. Nowadays we observe a process of depletion of many city centers, associated to the emergence of remote work. These trends have been amplified by the COVID-19 pandemic, and many city centers are now cheaper than the first suburban rings \cite{zillow} as it can be seen in Fig. \ref{D} b). However, the usual house pricing distribution in a circular framework is the one with a high-priced city center, and cheaper zones as we move away from it, i.e the core type represented in Fig. \ref{D} c). Both frameworks can be described via the Gaussian function:

\begin{equation}
  D(i_x,i_y)={A_{s,c}}\exp \left( {-\dfrac{(i_x-X_c)^2+(i_y-Y_c)^2}{4N}} \right) +B_{s,c},
  \label{eq:6}
\end{equation}
where $i_x,i_y=1,2...N$ being $N$ the system size. The point $(X_c,Y_c)$ corresponds to the center of the city, $X_c=Y_c=(N+1)/2$. Subindices $s$ and $c$ denote the suburban and core models, respectively. For the suburban model  $A_{s}=-2.507$ and $B_{s}=-0.525$ yielding a house pricing $D(i_x,i_y)$ in the range $[-3.025,-0.525]$. For the core model  $A_{c}=2.507$ and $B_{c}=-1.75$ give rise to a house pricing in the interval $[-1.75,0.75]$ as it shown in Fig. \ref{D} c).

Finally, in the grid city scheme we add some internal boundaries to the model. Towns are not continuous structures: roads, railways, highways, etc. tend to divide the city in order to facilitate the communication between its parts. The rectangular grid has been in use since Roman times \cite{stan_1946}, and is characterized by rectangular blocks and streets which meet at right angles. However, some areas created in this grid are more valuables than others, generally speaking those close to the city center. Therefore, we considered a high-priced center and also a rectangular grid with internal boundaries where the agents can not be relocated into, as it is illustrated in Fig. \ref{D} d).  

%
\section{Results and Discussion}
\label{sec:results}
As we can see in Eq. \ref{eq:2} the model parameters are the tolerance level \emph{T}, house pricing distribution $D_i$  and half the financial gap between the two social groups \emph{H}. 

Analyzing the segregation maps \cite{segregationUSA} of cities where segregation is known to happen, such as Detroit, Memphis, Chicago, etc. is easy to conclude that tolerance must be low. Thus, we consider $T=0.25$ for all the agents. This value means that an agent does not consider relocation to another spot if three or more of the new neighbors are different, unless the new economic environment is very favorable as it was explained in Sec. \ref{sec:model}.

The considered values of $H$ are $0,0.125$ and $0.250$. Notice that $H=0$ implies economic equality between the two social groups. The house pricing distribution $D_i$ depends only on the city type.

The system dynamics always tends to a stationary state \cite{ortega_g}. Although it evolves quickly, our condition for equilibrium is that the agent distribution remains constant for a long period of time, i.e. at least 50 Monte-Carlo (MC) steps.   Each result from the following sections, unless otherwise stated, is obtained for $500$ simulations, being the system size $N=50$. 

The next subsections aim to characterize the clusters inhabited by the segregated minority and how changes in the model parameters affect them. We study their size distribution in \ref{subsec:SCS}, the relation between its size and the location of the center of mass in \ref{subsec:PSC}, and finally their evolution in \ref{subsec:CE}.

\subsection{Segregated Cluster Size}
\label{subsec:SCS}
When $H\neq0$ red agents enjoy a favorable environment while blue agents are economically handicapped and tend to group into smaller clusters, i.e. ghettos. The cluster sizes $s$ are defined by their total number of blue agents. However, as our lattice is finite, we have included an upper cut-off size, $s_{max}$, which improves the parameter estimation \cite{winsize_2014}. Hence, we have obtained histograms of $s$ and fitted them to a probability density function (PDF) of the form $p(s)=Cs^{\alpha}$ where $C$ is the normalization constant and $\alpha <0$ accounts for the scaling exponent. For reasons of numerical stability \cite{clauset_2009} we focus on its complementary cumulative distribution function (CCDF), $P(s)$, that can be written as:

\begin{equation}
  CCDF=P(s)=\text{Pr}(X>s)= \int_{s}^{s_{max}}Cx^{\alpha}dx=\dfrac{C}{\alpha+1}(s_{max}^{\alpha+1}-x^{\alpha+1}) \quad \textrm{with} \quad \alpha  \neq{-1}.
  \label{eq:7}
\end{equation}

\begin{figure}[H]
\centering
\begin{tabular}{ccc}
\includegraphics[width=5.25cm, height=4.2cm]{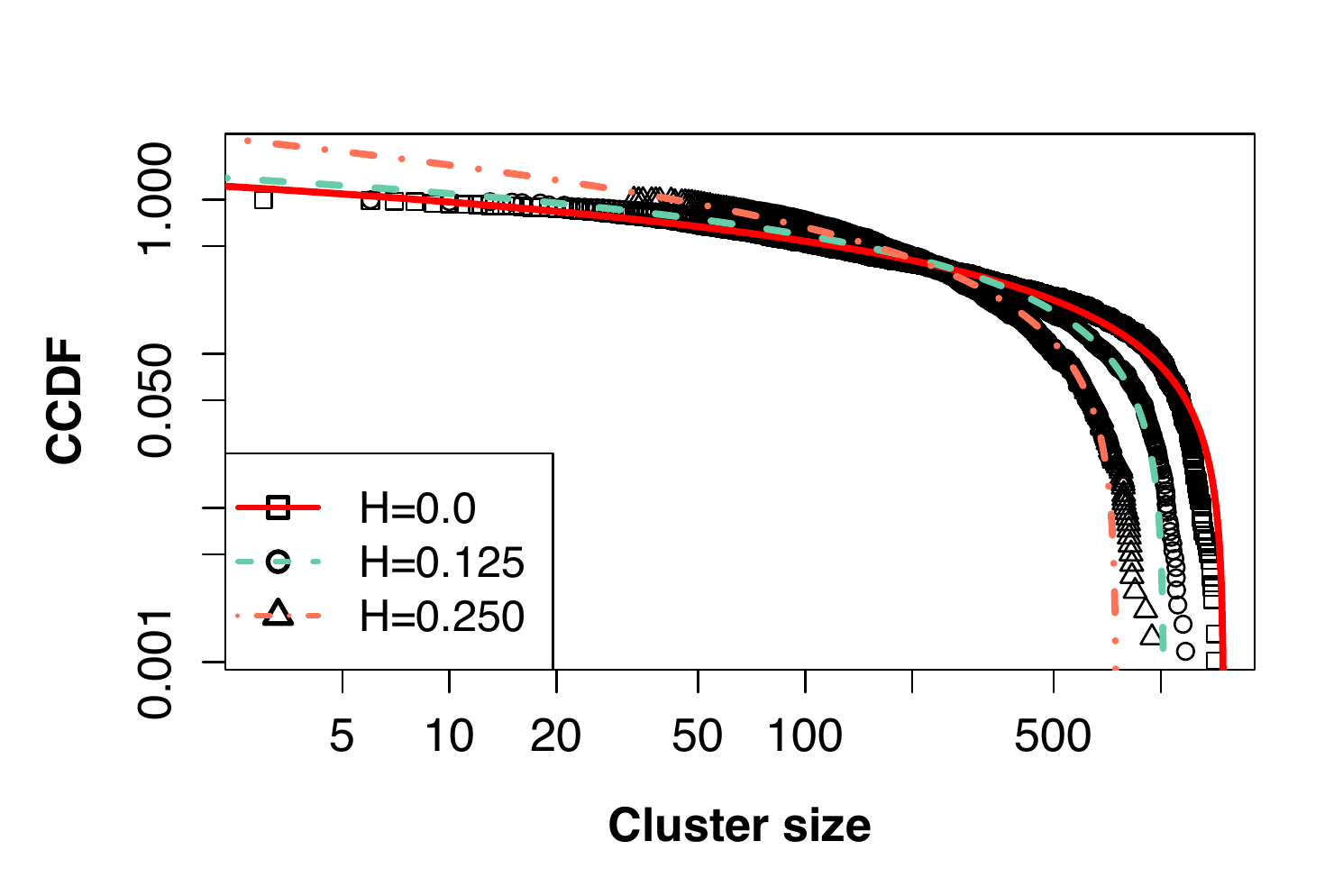}&
\includegraphics[width=5.25cm, height=4.2cm]{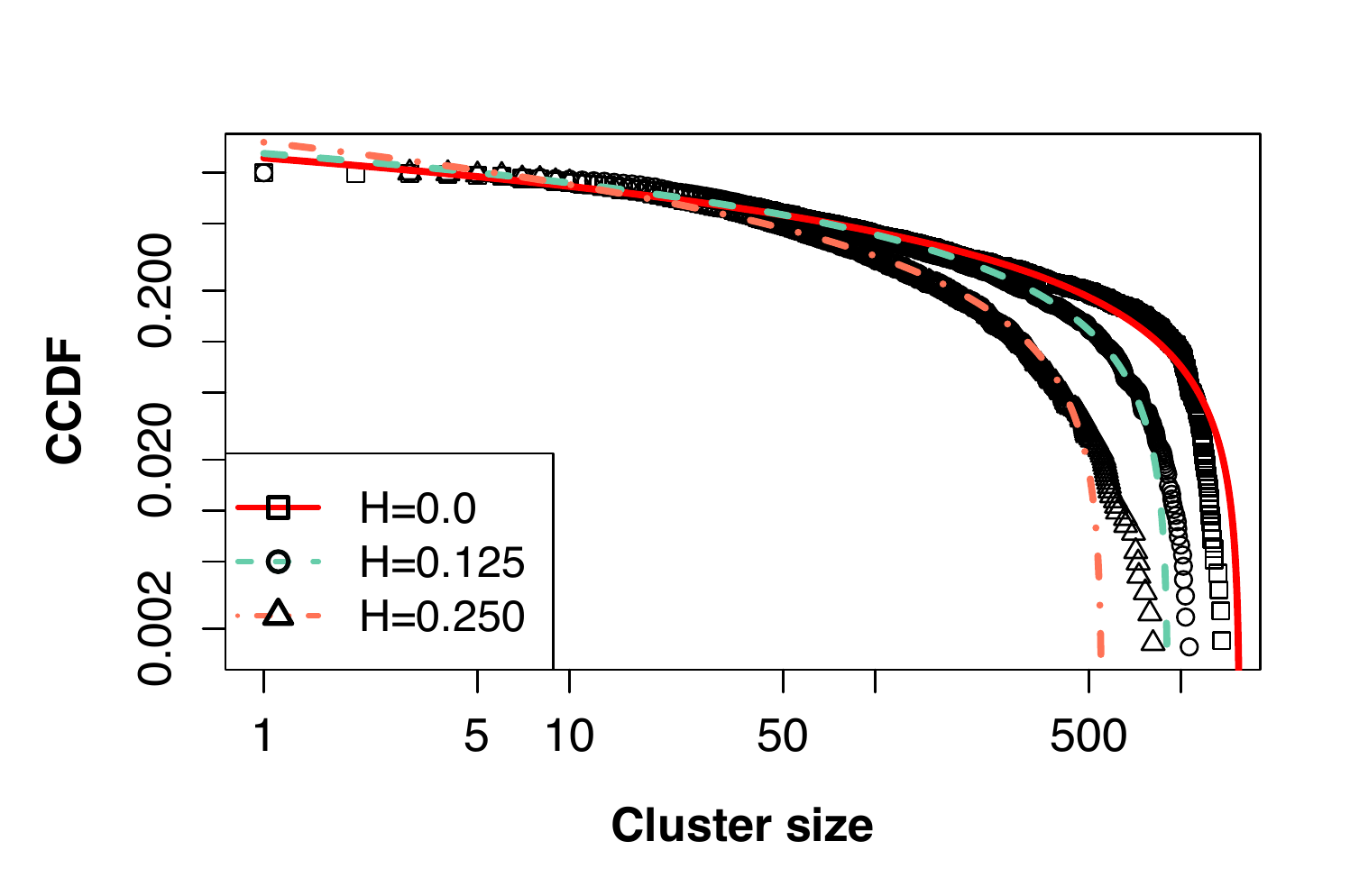}&
\includegraphics[width=5.25cm, height=4.2cm]{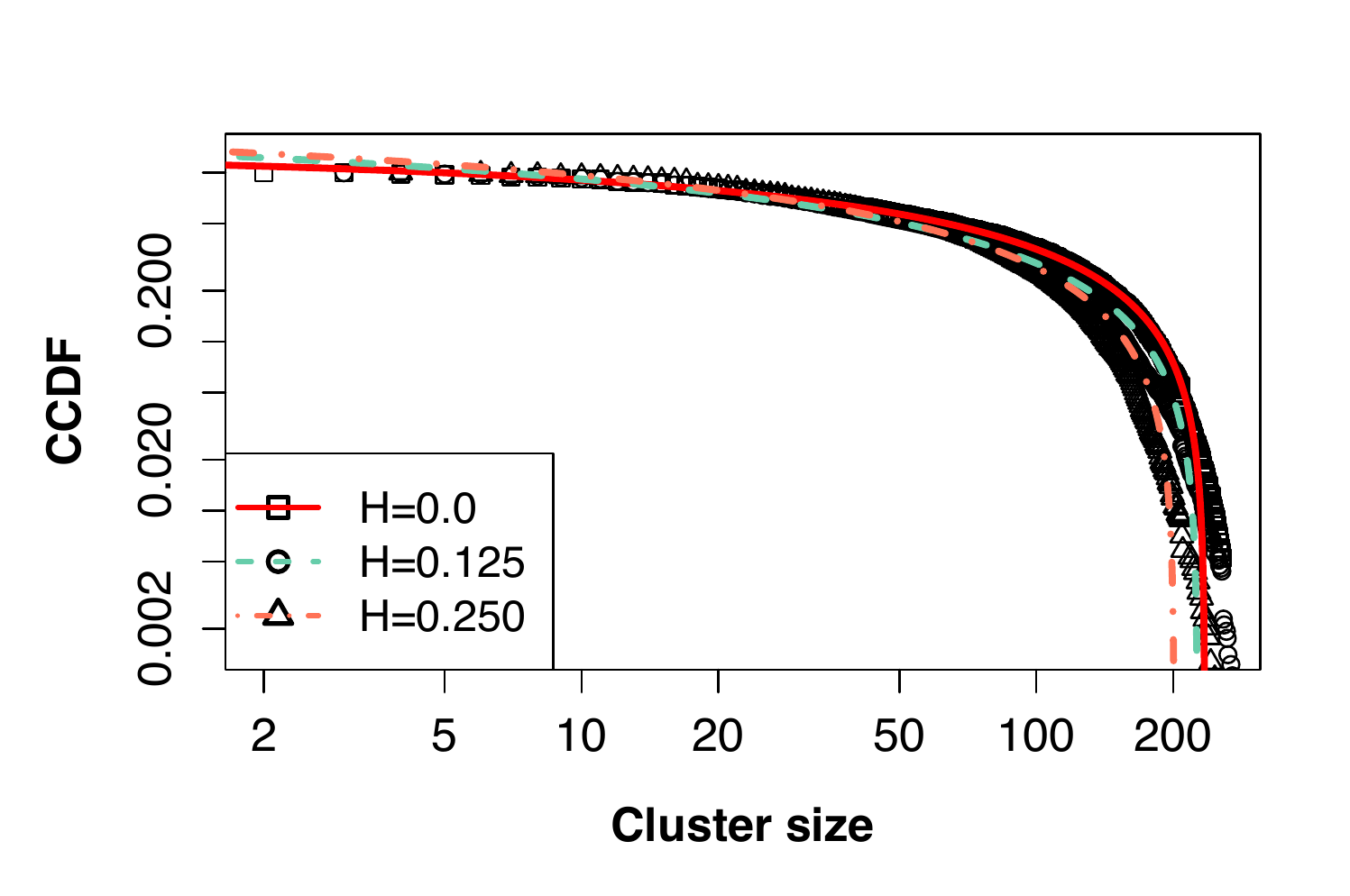}\\
a)&b)&c)\\
\end{tabular}
\caption{Complementary cumulative distribution function (CCDF) of the blue cluster sizes for $T=0.25$. Letters denote the considered city type:   flat a) vertical b) and grid c) models. The fitted power-law functions are depicted with lines, being its $H$ value specified. }
\label{fig3aval}
\end{figure}

Blue cluster expansion in the city is constrained by several factors: the extension of the zone that is not in the vacancy dominated regime, red cluster expansion, system borders and internal boundaries if considered. We have computed the adjusted r-squared coefficient,  $r_{adj}^2$ as a goodness-of-fit measure via the Wherry formula  \cite{rsquared}. As it can be seen in Fig. \ref{fig3aval} this approximation shows good agreement to our data, being $r_{adj}^2>0.97$ for all the fitted lines.

The effect of $H$ over the cluster size can be anticipated from Eq. \eqref{eq:2} and it is illustrated in Fig. \ref{fig3aval}. As $H$ increases the city becomes progressively unaffordable for the blue agents, making them feel unsatisfied and leave the lattice. 

Now, we focus our attention on the fitted parameters from Eq. \eqref{eq:7}: the power law exponent $\alpha$ and the upper cut-off size $s_{max}$. On one hand $\alpha$ is located in the narrow range $[-1.135,-0.69]$ suggesting a scale independent of $H$. The values shown in Table \ref{table0} are typical of processes associated with gentrification phenomena \citep{ortega_g}. The interval of $\alpha$ for the grid model is $[-0.82,-0.69]$ highlighting the edges influence.

\begin{table}[H]
\begin{centering}
\begin{tabular}{|c|c|c|c|}
\hline
$\alpha$&\textbf{$H=0$}  & \textbf{$H=0.125$} & \textbf{$H=0.250$ }\\
\hline 
\hline 
\textit{Flat} & $-0.974\pm 0.005$ &$-0.941\pm 0.005$ & $-1.135\pm 0.007$ \\
\hline 
\textit{Vertical} & $-1.012\pm 0.005$ &$-1.020\pm 0.006$ & $-1.121\pm 0.006$ \\
\hline 
\textit{Suburban}& $-1.049\pm 0.003$ &$-1.090\pm 0.003$ & $-0.696\pm 0.009$ \\
\hline 
\textit{Core} & $-0.875\pm 0.003$ &$-0.9131\pm 0.009$ & $-0.97\pm 0.02$ \\
\hline 
\textit{Grid} & $-0.69\pm 0.01$ &$-0.82\pm 0.02$ & $-0.82\pm 0.02$ \\
\hline 
\end{tabular}

\par\end{centering}
\caption{Values of the power law exponent  $\alpha$ for the considered city types. }
\label{table0}
\end{table}

On the other hand $s_{max}$, which is related to the maximum cluster size, shows a strong dependence on the economic gap as it can be observed in Table \ref{table1}.  When $H$ increases from $0$ to $0.250$, $s_{max}$ is reduced approximately to half its value except for the grid case: 

\begin{table}[H]
\begin{centering}
\begin{tabular}{|c|c|c|c|}
\hline
$s_{max}$&\textbf{$H=0$}  & \textbf{$H=0.125$} & \textbf{$H=0.250$ }\\
\hline 
\hline 
\textit{Flat} & $1501\pm 10$ &$1019\pm 6$ & $748\pm 5$ \\
\hline 
\textit{Vertical} & $1510\pm 20$ &$889\pm 9$ & $577\pm 6$ \\
\hline 
\textit{Suburban}& $1453\pm 8$ &$1064\pm 7$ & $641\pm 5$ \\
\hline 
\textit{Core} & $1256\pm 7$ &$719\pm 7$ & $490\pm 6$ \\
\hline 
\textit{Grid} & $235\pm 1$ &$226\pm 2$ & $201\pm 2$ \\
\hline 
\end{tabular}
\par\end{centering}
\caption{Upper cut-off cluster size, $s_{max}$ for different city types.}
\label{table1}
\end{table}

Large segregated clusters are present in flat, vertical and suburban cities as we can observe in Table \ref{table1}. Nevertheless, core and grid models exhibit lower values of the maximum size. This implies that red agents and the geometrical distribution of $D_{i}$ may act as a constraint on ghetto growth in these city frameworks, reducing $s_{max}$ as the data from Table \ref{table1} shows. The grid model is greatly affected by the existence of these internal borders. In fact, as the internal boundaries divide the system into nine blocks with $16\times16$ cells, cluster sizes contain up to $256$ agents. The social meaning is clear, the borders prevent both gentrification and blue cluster expansion.

\subsection{Segregated clusters location}
\label{subsec:PSC}
 
In order to characterize the ghetto locations the coordinate axis and the system geometry must be defined. Their choice depends on the considered city type. For the flat and vertical frameworks we only consider the average distance to the top of the lattice, $i_y$. From here on, distances will be normalized dividing the measured lengths by the system size $N$.

\begin{figure}[H]
\centering
\begin{tabular}{ccc}
\includegraphics[width=5.25cm, height=4.2cm]{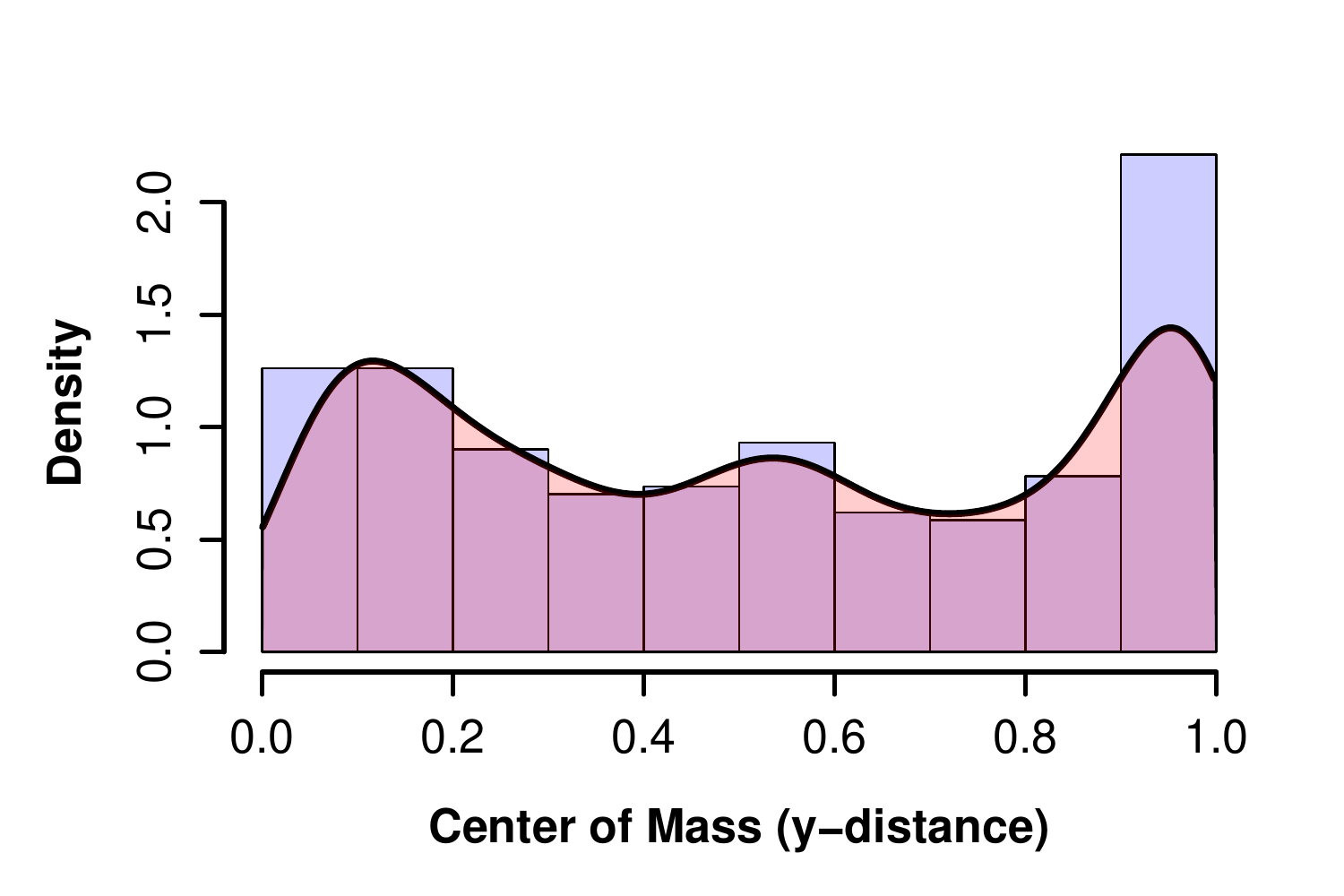}&
\includegraphics[width=5.25cm, height=4.2cm]{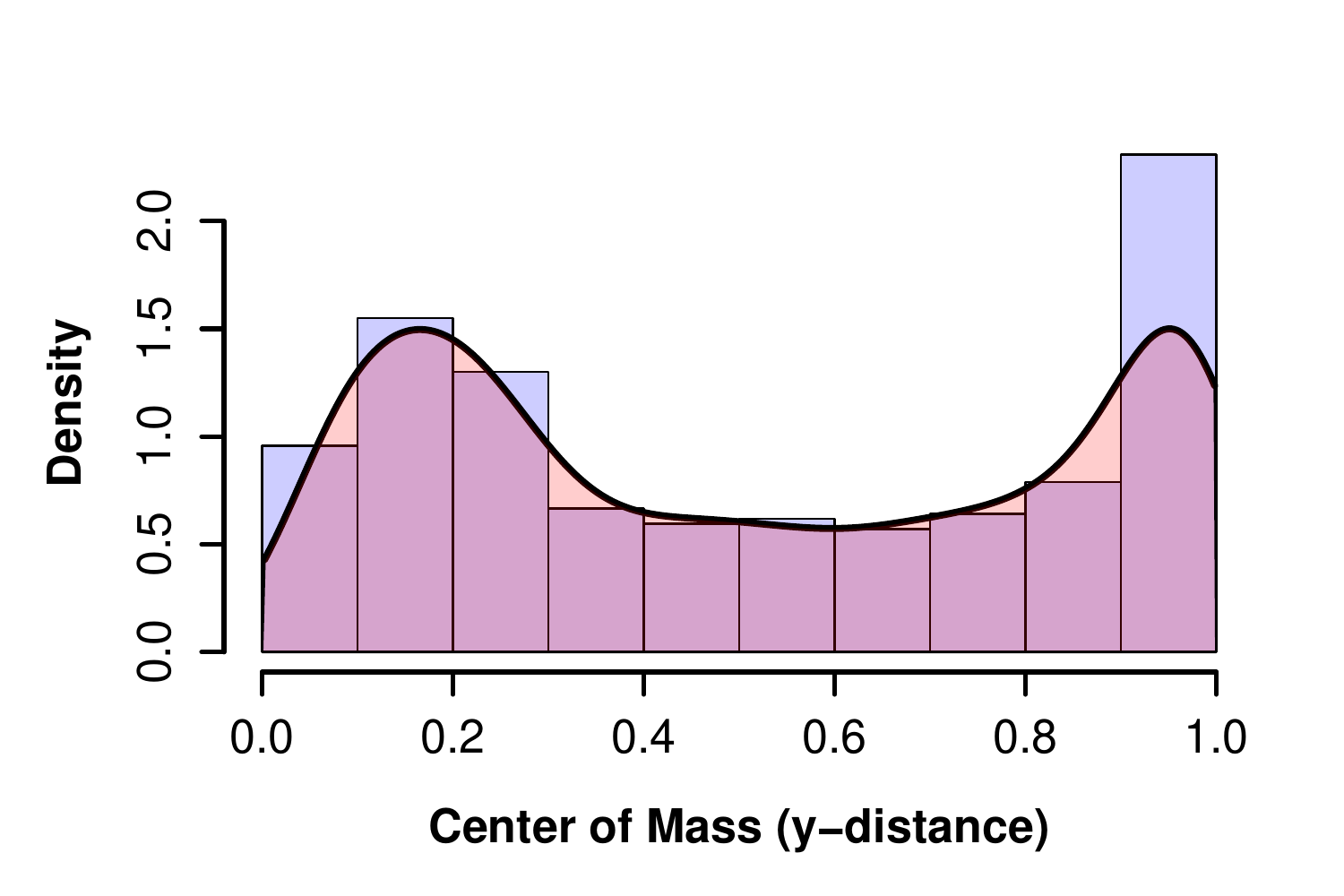}&
\includegraphics[width=5.25cm, height=4.2cm]{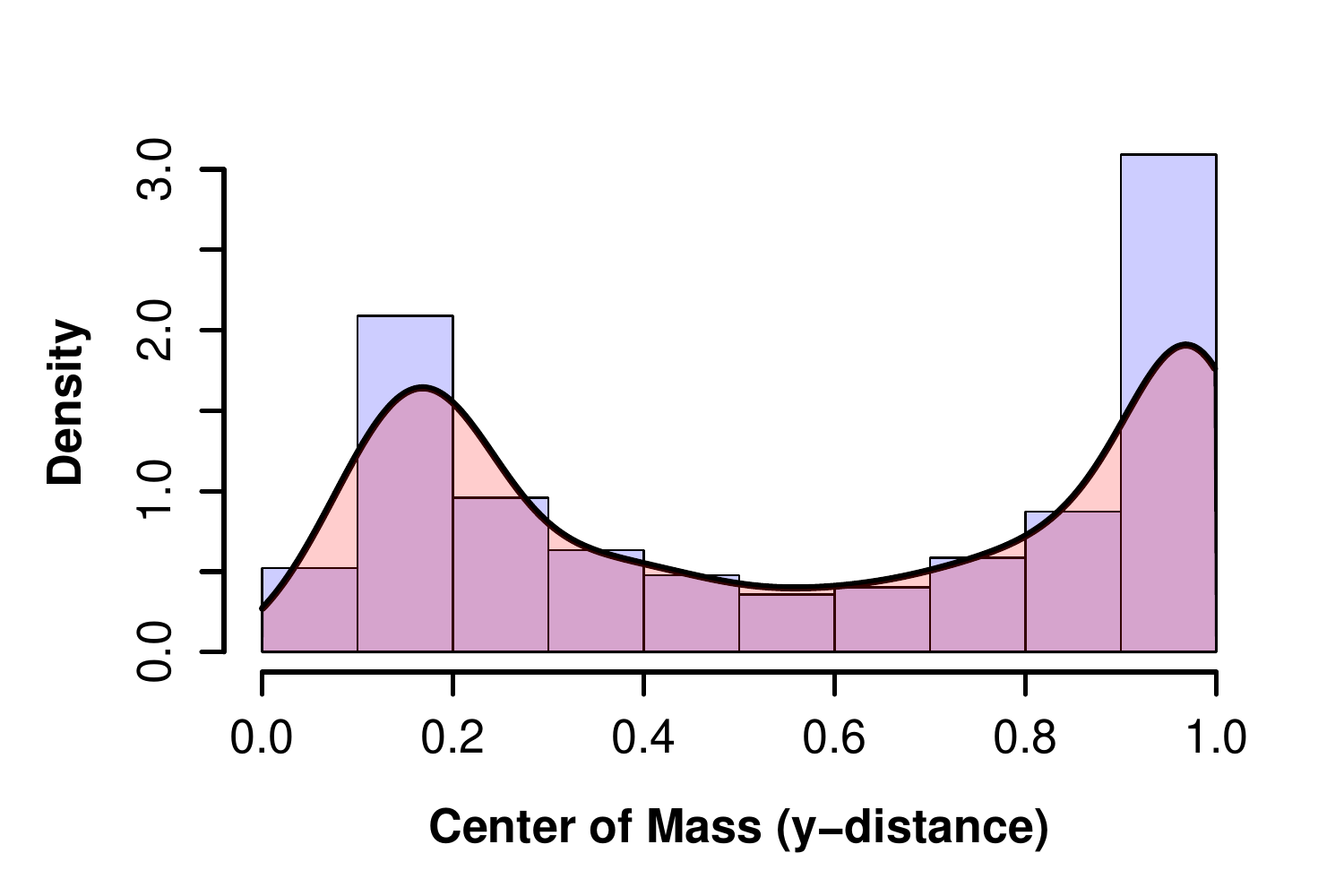}\\
a)&b)&c)\\
\end{tabular}
\caption{Density histograms of center of mass positions for the vertical model with $H=0$ a) $H=0.125$ b) and $H=0.250$ c). Distances are measured from the top of the lattice and normalized via the system size $N$. }
\label{4hist}
\end{figure}

We focus our attention on the vertical framework. Density histograms for this city type and different $H$ values are presented in Fig. \ref{4hist}. We must note the presence of two peaks near the extremes of the considered interval. This accumulation is due to the border effect: close to the system edges the number of neighbors is reduced, thus decreasing the effects of segregation terms over happiness in Eq. \eqref{eq:2}. Hence, ghettos can be established near the city limits.  Besides, another interesting phenomenon can be inferred from Fig. \ref{4hist}. As the economic gap increases the densities of segregated neighborhoods in the city center are reduced, and ghettos are displaced towards the city borders. In Fig. \ref{4hist} c) more ghettos are established for further lengths which correspond to the city most affordable area.  

\begin{figure}[H]
\centering
\begin{tabular}{ccc}
\includegraphics[width=5.25cm, height=4.2cm]{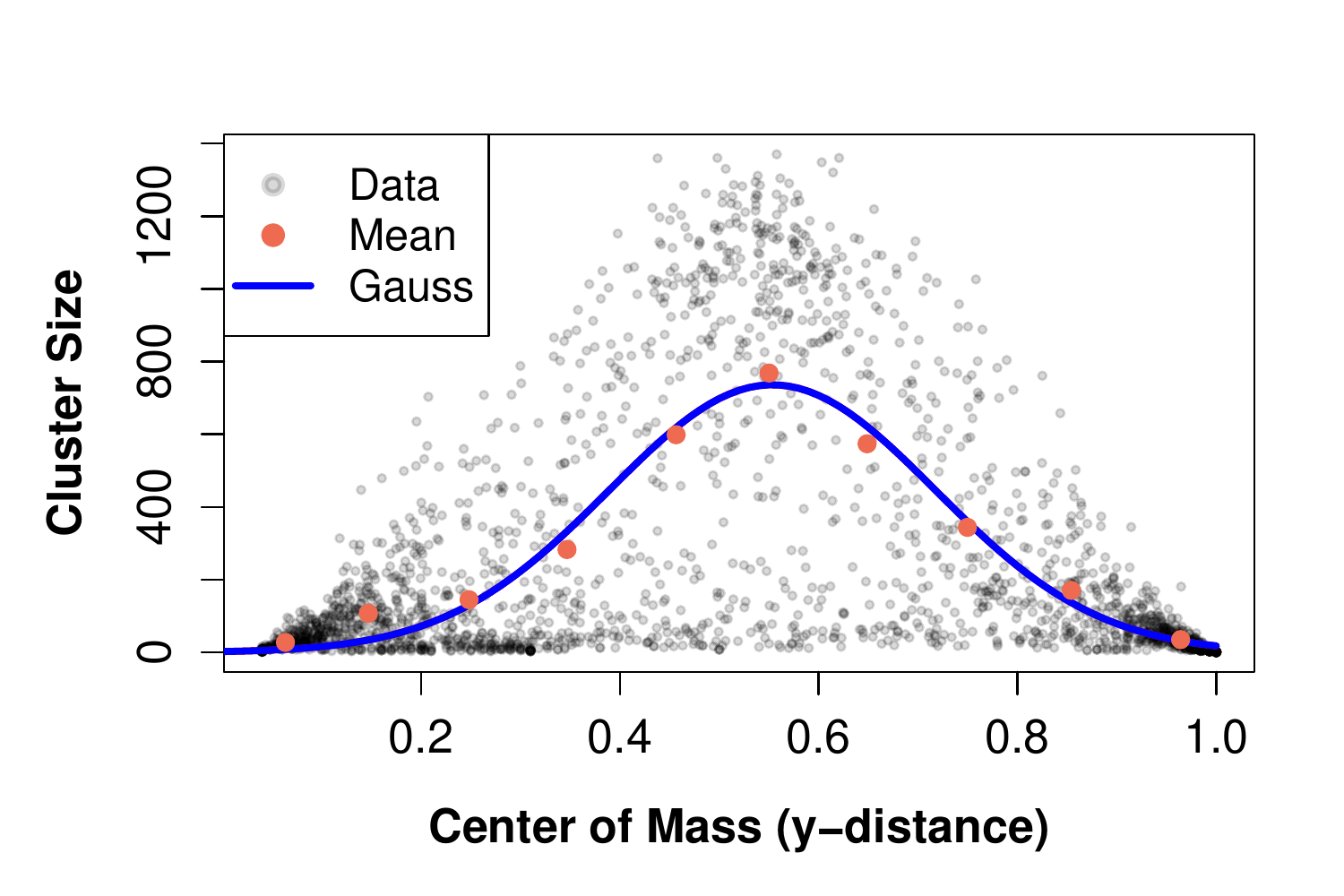}&
\includegraphics[width=5.25cm, height=4.2cm]{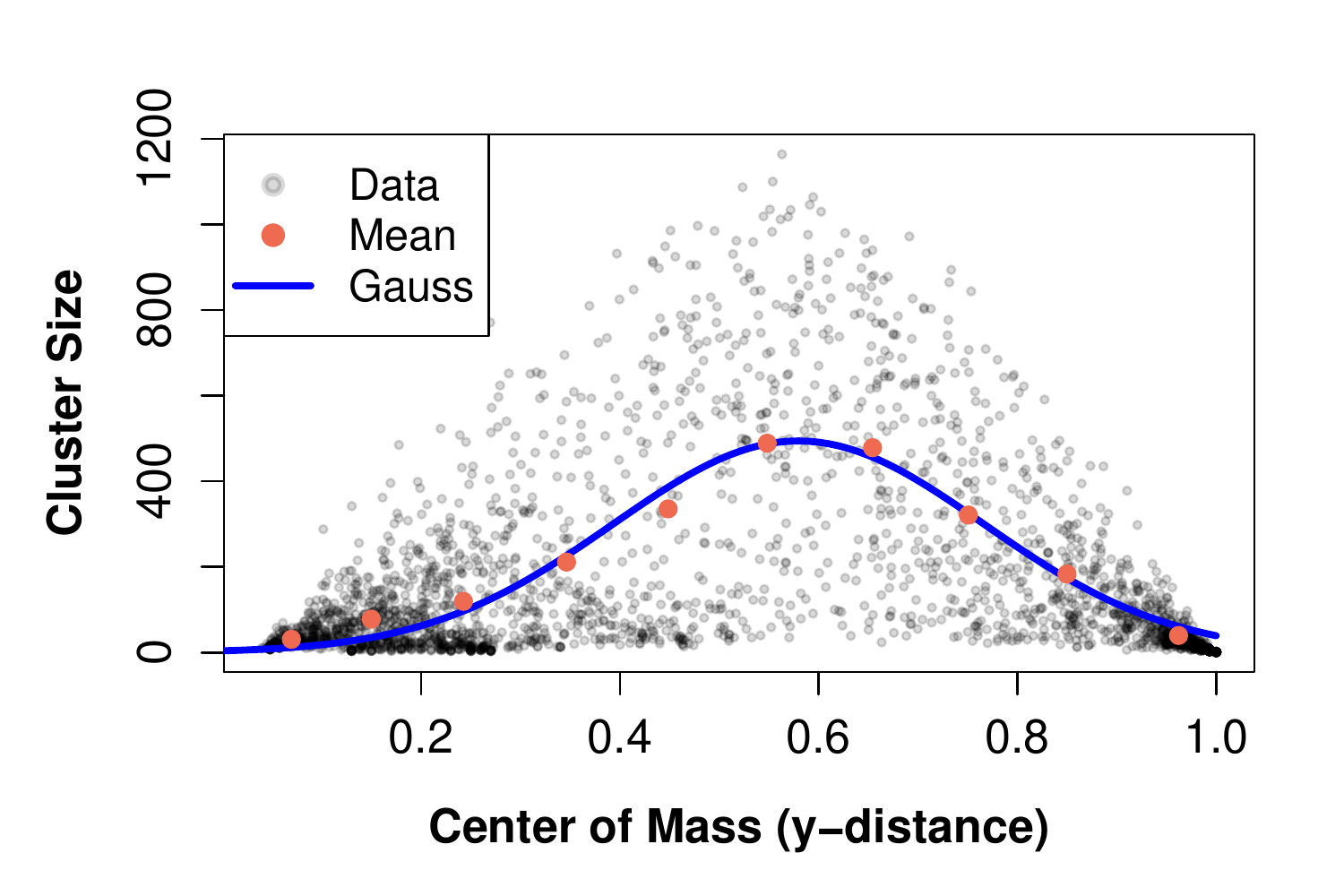}&
\includegraphics[width=5.25cm, height=4.2cm]{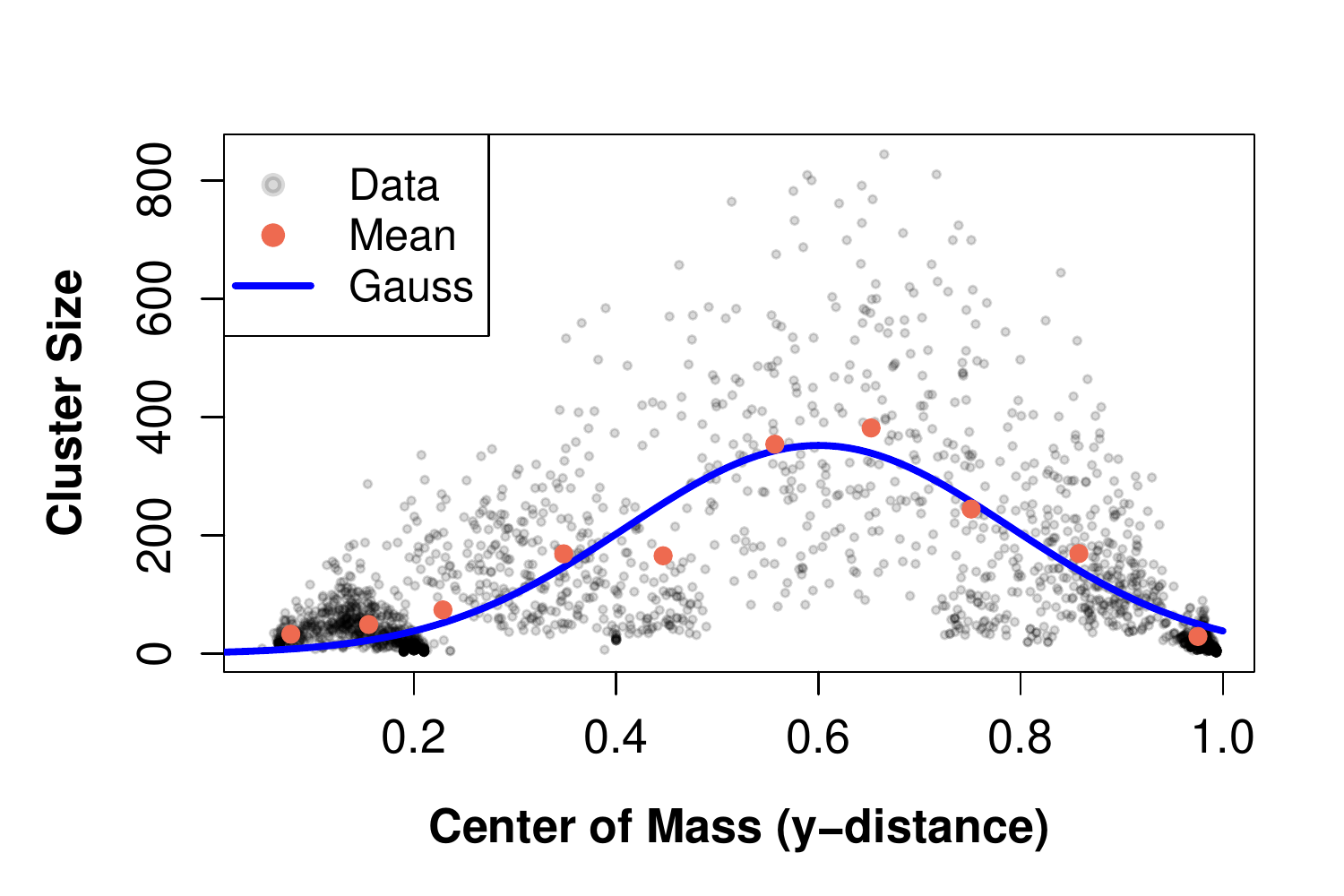}\\
a)&b)&c)\\
\end{tabular}
\caption{Cluster size and center of mass locations for the vertical model with $H=0$ a) $H=0.125$ b) and $H=0.250$ c).  The Mean points are calculated assigning intervals of $0.1$ width to the x-axis and averaging the position of all the points that lie in each range. The line labeled as Gaussian represents the fit of a Gaussian function to the data points. As the data show great variability values of $r_{adj}^2$ for the fit remain low. In a), b) and c) cases are, $0.53, 0.46$ and $0.51$, respectively. Distances are measured from the top of the lattice and normalized via the system size $N$. }
\label{5Gauss}
\end{figure}

We employed a scatter plot to represent the relation between the cluster sizes and the positions of the center of mass. The cases for the vertical model with $H=0, 0.125$ and $H=0.250$ are illustrated in Fig. \ref{5Gauss} a), b) and c), respectively.  

The average points in Fig. \ref{5Gauss} are calculated taking intervals of $0.1$ width in the x-axis and averaging the position of all the points that lies in each range. We also expect a functional dependence of the average cluster size with the position of the center of mass. We have fitted our experimental results to a Gaussian form,

\begin{equation}
  C\exp \left[ -\dfrac{(x-\mu)^2}{2\sigma^2}\right],
  \label{eq:8}
\end{equation}
where $C$ is an amplitude constant, $\mu$ the mean and $\sigma$ the standard deviation. This expression points at the existence of an optimal zone for the growth of blue neighborhoods, located by $\mu$, which we define as the \textit{optimal ghetto location}. This is the zone where most ghettos can achieve their maximum size. For the vertical model is close to $0.6$ as we can see in Fig. \ref{5Gauss}. However, as it can be deduced from histograms in Fig. \ref{4hist} clusters do not have to be created preferentially in this location. 

Now we shift our focus to radial cities. In these frameworks is useful to define the distance from the center of mass of the cluster to the center of the lattice, $R_{c}$:  

\begin{equation}
  R_{c}=\sqrt{\(\dfrac{1}{n_{tc}} \sum_b i_{b} - X_c \)^2+\(\dfrac{1}{n_{tc}} \sum_b j_{b} -Y_c \)^2}.
  \label{eq:9}
\end{equation}

In the previous equation $b$ is the index that runs over all the blue agents in the same cluster, the total number of agents in the cluster is given by $n_{tc}=\sum_{b} n_{b}$, while $i_{c}$ and $j_{c}$ are the $(x,y)$ integer coordinates of the considered cell. $X_c=Y_c=(N+1)/2$ are the center coordinates, being $N$ the system size. In radial models an optimal ghetto location also appears. Therefore, we have fitted the data to the  Gaussian curve from Eq. \eqref{eq:8}.

\begin{figure}[H]
\centering
\begin{tabular}{ccc}
\includegraphics[width=5.25cm, height=4.2cm]{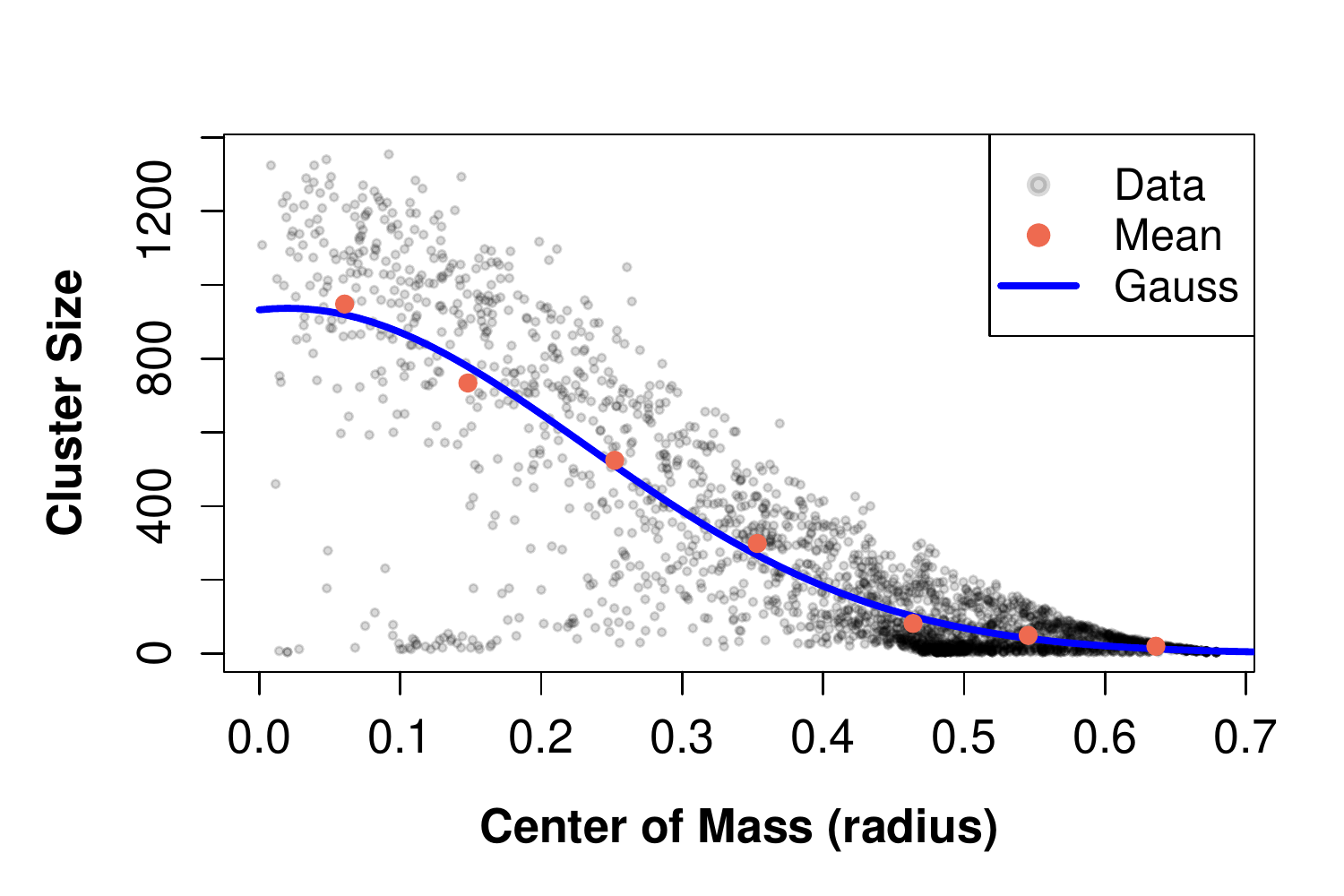}&
\includegraphics[width=5.25cm, height=4.2cm]{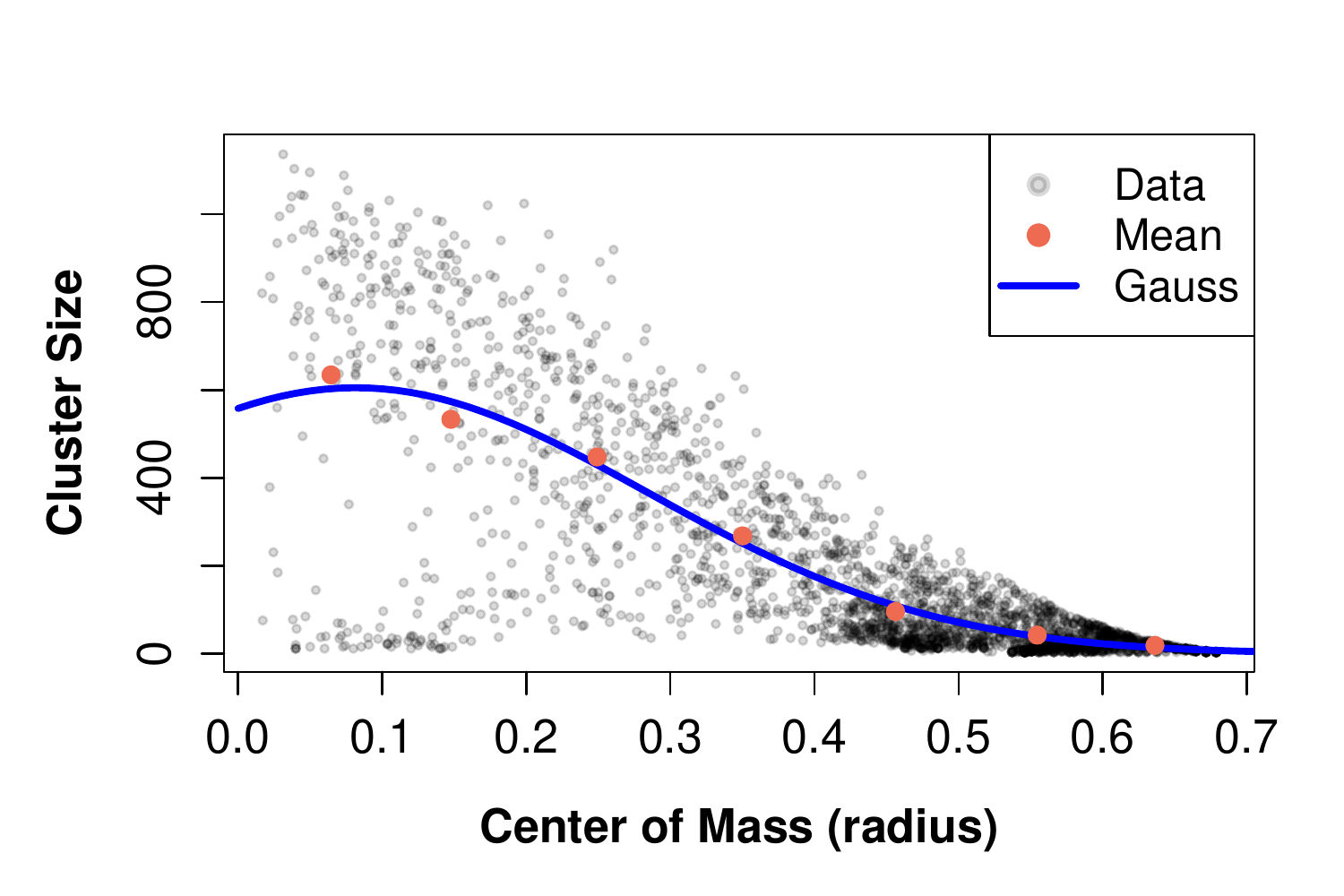}&
\includegraphics[width=5.25cm, height=4.2cm]{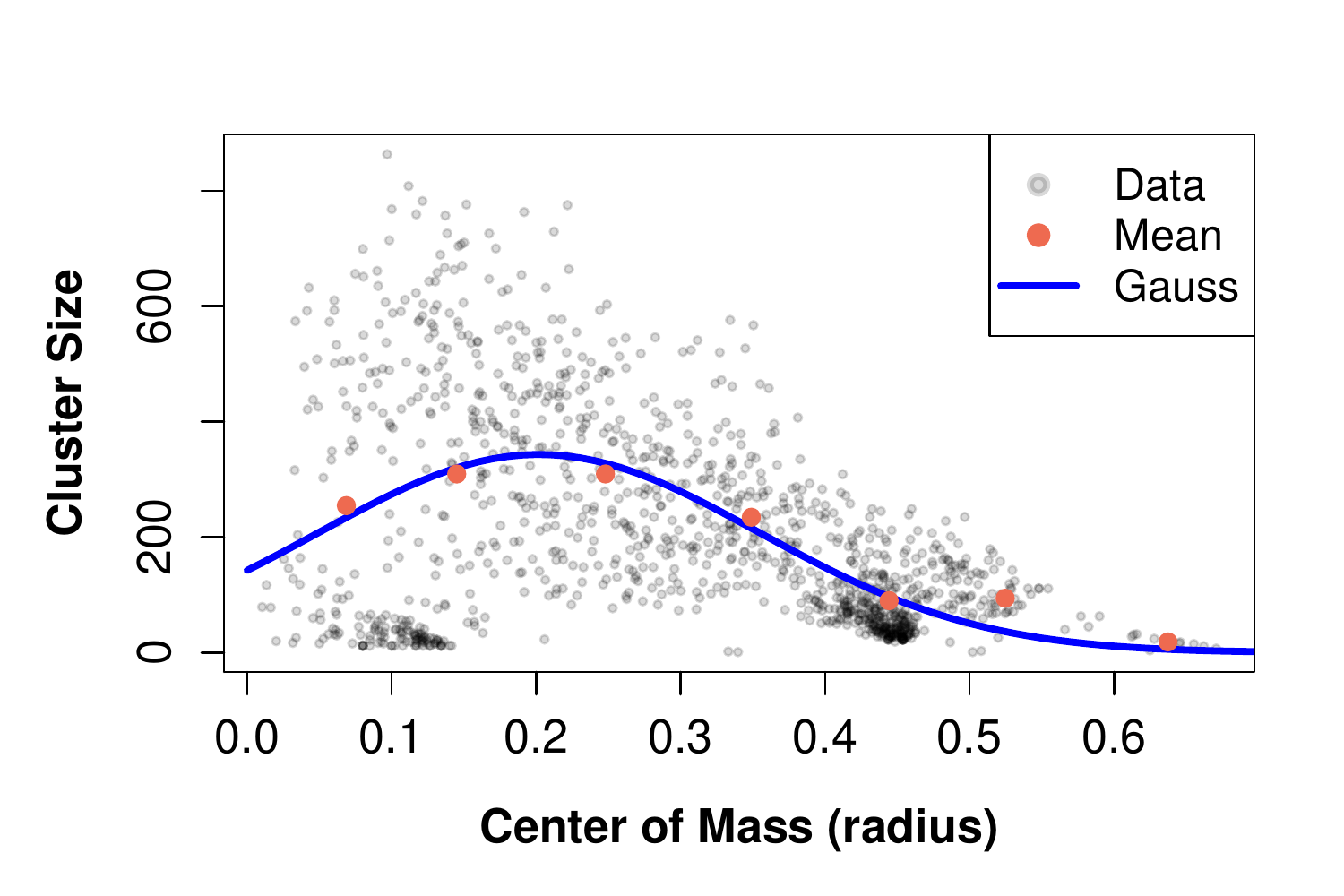}\\
a)&b)&c)\\
\end{tabular}
\caption{Cluster size and center of mass positions for the suburban model with $H=0, 0.125$ and $0.250$ in a), b) and c), respectively. Values of $r_{adj}^2$ are, $0.75, 0.64$ and $0.31$, in the previous order. The average points are calculated assigning intervals of $0.1$ width to the x-axis and averaging the position of all the points that lies in each range. The line labeled as Gaussian represents the fit to a Gaussian function.  Distances are measured from the center of the lattice and normalized by $N$. }
\label{6exp}
\end{figure}

The relation between the center of mass locations and the cluster size for the suburban model is depicted in Fig. \ref{6exp}.  In this framework both the optimal geometrical position and the most interesting economic zone coincide.  As it is depicted in Fig. \ref{6exp} a) cluster sizes decrease as the radial distance increases, due to the system edges which restrict the cluster expansion. Consequently, the optimal cluster location is close to the origin (Table \ref{table2}). We can also observe that the optimal ghetto location is being progressively displaced from the center as $H$ increases, as it is depicted in Fig. \ref{6exp} b) and c). This can be explained via the red agents expansion, i.e. as they can also expand from the borders to the center, blue agents are displaced from the central zone. Thus, dark zones for a radius greater than $0.4$ in Fig. \ref{6exp} highlight that most ghettos are created far from the center and their growth is constrained by the system edges.

After the two previous detailed examples we discuss the optimal ghetto locations for the different type of cities by means the values from Table \ref{table2}. From a geometric point of view, the best zone for a cluster to expand is the center of the city. This fact is highlighted by the flat city model. The lattice middle point expressed in normalized units is located at the center $\approx 1/2$. As it can be observed in Table \ref{table2} larger clusters are formed in the city center independently of the value of $H$. In contrast, for the vertical model,  the agents can find better economic perspectives at larger values of $y$. In addition to this, when $H$ increases blue agents suffer economic deprivation. As a consequence they need to find better economic areas, thus displacing the optimal ghetto location, $\mu$, to larger $y$ distances, as it can be seen in the normalized values from Table \ref{table2}. In other words, the interplay between geometric and economic factors affects the optimal ghetto location. 
 
\begin{table}[H]
\begin{centering}
\begin{tabular}{|c|c|c|c|}
\hline
$\mu$&\textbf{$H=0$}  & \textbf{$H=0.125$} & \textbf{$H=0.250$ }\\
\hline 
\hline 
\textit{Flat} & $0.512\pm 0.005$ &$0.512\pm 0.006$ & $0.511\pm 0.007$ \\
\hline 
\textit{Vertical} & $0.553\pm 0.004$ &$0.580\pm 0.004$ & $0.600\pm 0.004$ \\
\hline 
\textit{Suburban}& $0.02\pm 0.01$ &$0.08\pm 0.01$ & $0.202\pm 0.006$ \\
\hline 
\textit{Core} & $0.00\pm 0.03$ &$0.267\pm 0.008$ & $0.332\pm 0.004$ \\
\hline 
\end{tabular}

\par\end{centering}
\caption{Normalized distances to the optimal ghetto locations, $\mu$, for different city types.  For the flat and vertical models distance is measured from the top of the city. In radial cases, lengths are estimated from the lattice middle point. Distances are normalized via $N$. }
\label{table2}
\end{table}

An interesting phenomenon arises in the core framework. When $H=0$ agents are economically similar and large clusters are located in the city center.  However, when $H\neq0$, the expansion of red agents and economic issues force blue agents to move further away from the center than in the suburban case, as Table \ref{table2} shows. This result is mainly due to the combination of two factors: blue agents becoming more and more monetarily handicapped as we can deduce from Eq. \ref{eq:2}, where $H(i)=+H$, thus leaving the expensive center. In addition to this, the economic condition for red agents becomes more advantageous, $H(i)=-H$. Therefore, they tend to expand through the center and may force blue agents to move out, as it happened in the suburban model. Since the system edges impose a restriction on the cluster growth, a maximum is expected between the city center and the lattice borders.

\begin{figure}[H]
\centering
\begin{tabular}{cc}
\includegraphics[width=5cm, frame]{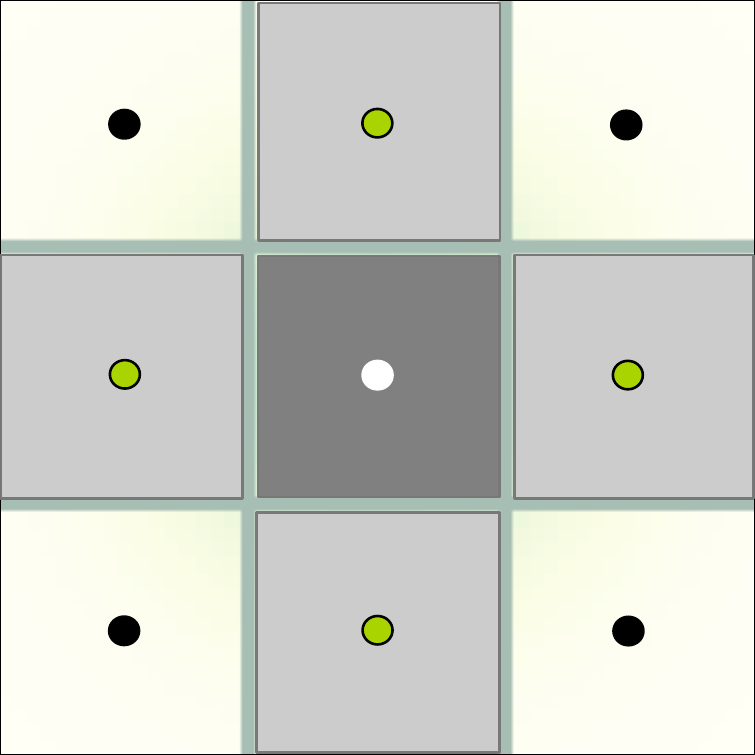}&
\includegraphics[width=7.5cm, frame]{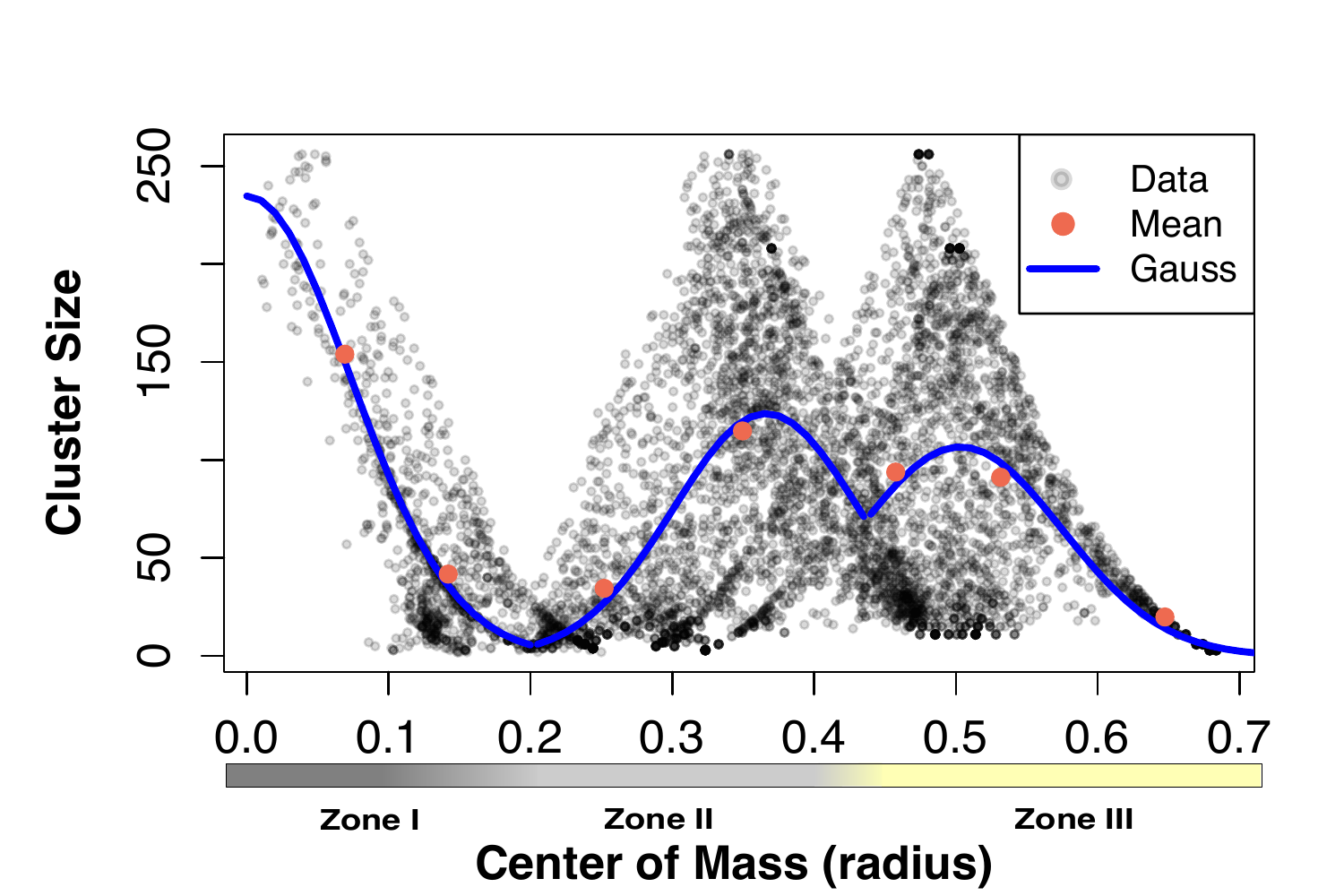}\\
a)&b\\
\end{tabular}
\caption{a) Scheme of the grid model considering their distance to the central point. Clearer tones are associated with larger distances to the origin. Points in each block are located on the center. b) Relation between center of mass radial distance and cluster size for $H=0$. Each colour under the x-axis is related to the zones scheme in a). The $r_{adj}^2$ values for fitted lines in zone I, II and III are $0.70$, $0.30$ and $0.18$, respectively.  These low values are dued to overlapping regions between zones. Graphs for $H=0.125$ and $H=0.250$  are not depicted given their similarity with Fig. \ref{7square} b).  }
\label{7square}
\end{figure}

Finally we analyze the grid model in Fig \ref{7square}. As it is illustrated in the scheme from Fig \ref{7square} a), internal borders divide the city into nine blocks: the dark one which is placed in the center, four gray blocks corresponding to the four cardinal directions and the clearer ones which are the furthest from the lattice middle point. As we focus our attention on the distance to the center we group the blocks into three zones: I, II and III, following the same previous order. Correspondingly we can appreciate the same color scheme used for the blocks associated to each zone under the x-axis of Fig. \ref{7square} b). This points out that major contributions to the fitted lines come from these areas. Nevertheless, there are transition areas with contributions from two zones which are characterized by faded tones in the Fig. \ref{7square} b) color bar.

Three peaks can be found in  Fig. \ref{7square} b), each one of them is located in a different region. As we have previously said, the appearance of a peak corresponds to the interplay between the geometrical constraints and the house pricing distribution. From the geometrical perspective, the city presents internal barriers. Thus, optimal positions to create large clusters are the center of each block, represented by points in Fig. \ref{7square} a) scheme. On the other hand the city central zone  is expensive. Thus a shift towards larger distances is also expected. The experimental and fitted values can be found in Table \ref{table3}.
 
\begin{table}[H]
\begin{centering}
\begin{tabular}{|c|c|c|c|c|}
\hline
\textit{$\mu$}&\textit{Fit}&\textbf{$H=0$}  & \textbf{$H=0.125$} & \textbf{$H=0.250$ } \\
\hline 
\hline 
\textit{Zone I} & $0$&$0\pm0.008$ &$0\pm0.002$ & $0.159\pm 0.002$ \\
\hline 
\textit{Zone II}& $0.34$& $0.367\pm 0.002$ &$0.366\pm 0.002$ & $0.364\pm 0.002$  \\
\hline 
\textit{Zone III} & $0.48$&$0.503\pm0.003$ &$0.498\pm 0.003$ & $0.514\pm 0.004$  \\
\hline 
\end{tabular}
\par\end{centering}
\caption{Optimal ghetto locations $\mu$ for the zones in Fig. \ref{7square}. Theoretical values correspond to the normalized distances from the lattice middle point, the white circle, to the center of each block represented by coloured spots in Fig. \ref{7square} a).  } 
\label{table3}
\end{table}

We should note the closeness between the theoretical and the experimental values in Table \ref{table3}. The interpretation is straightforward: barriers help segregated clusters to disminish the effect of the economic gap. Boundaries act as a barrier to the red clusters growth decreasing gentrification effects. This can also be understood as the creation of ghettos whose boundaries isolate them from the rest of the city favouring segregation \cite{blanchard}.

\subsection{Segregated cluster evolution}
\label{subsec:CE}

\begin{figure}[H]
\centering
\begin{tabular}{ccc}
\includegraphics[width=5.3cm, height=3.9cm]{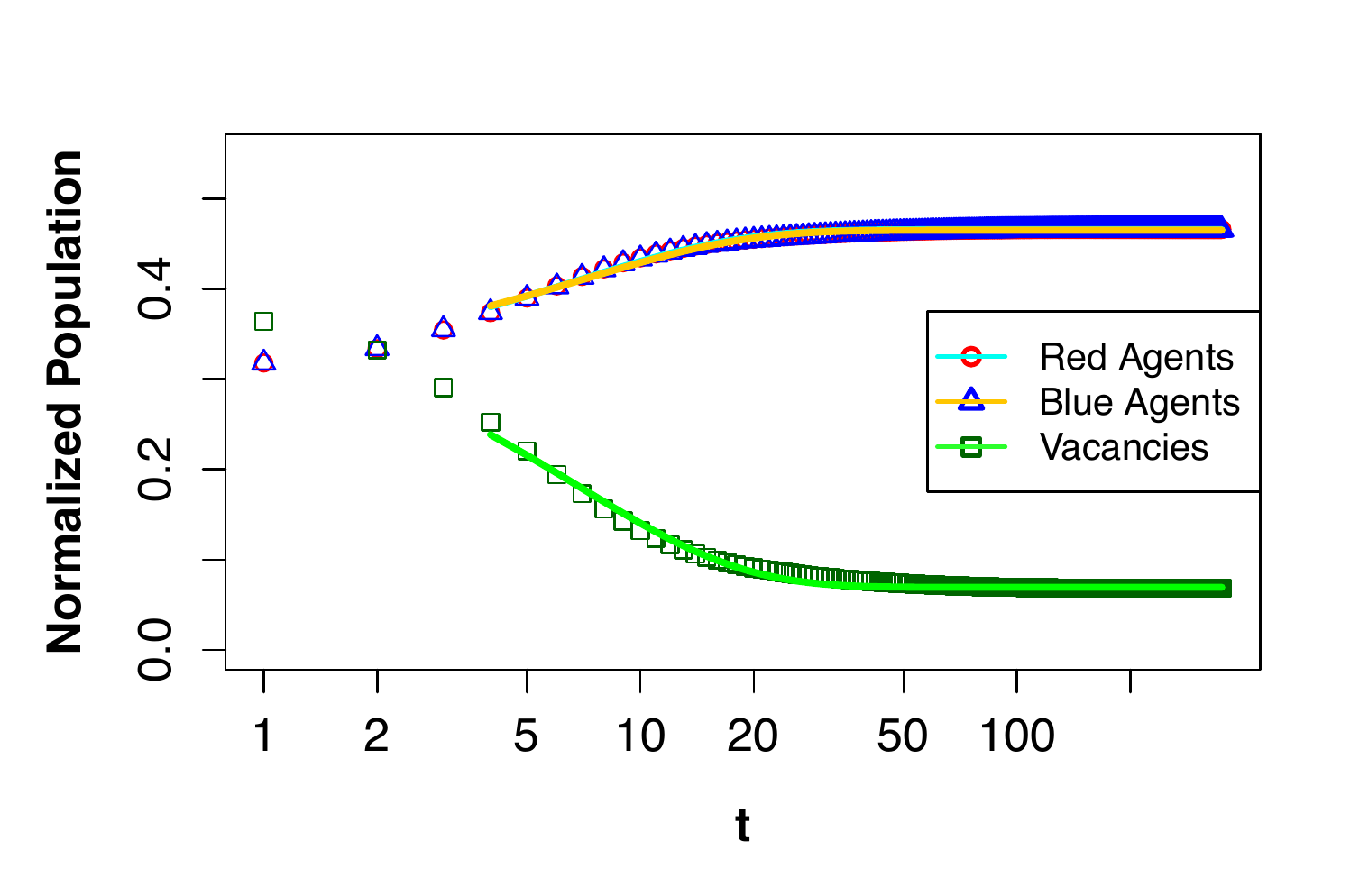}&
\includegraphics[width=5.3cm, height=3.9cm]{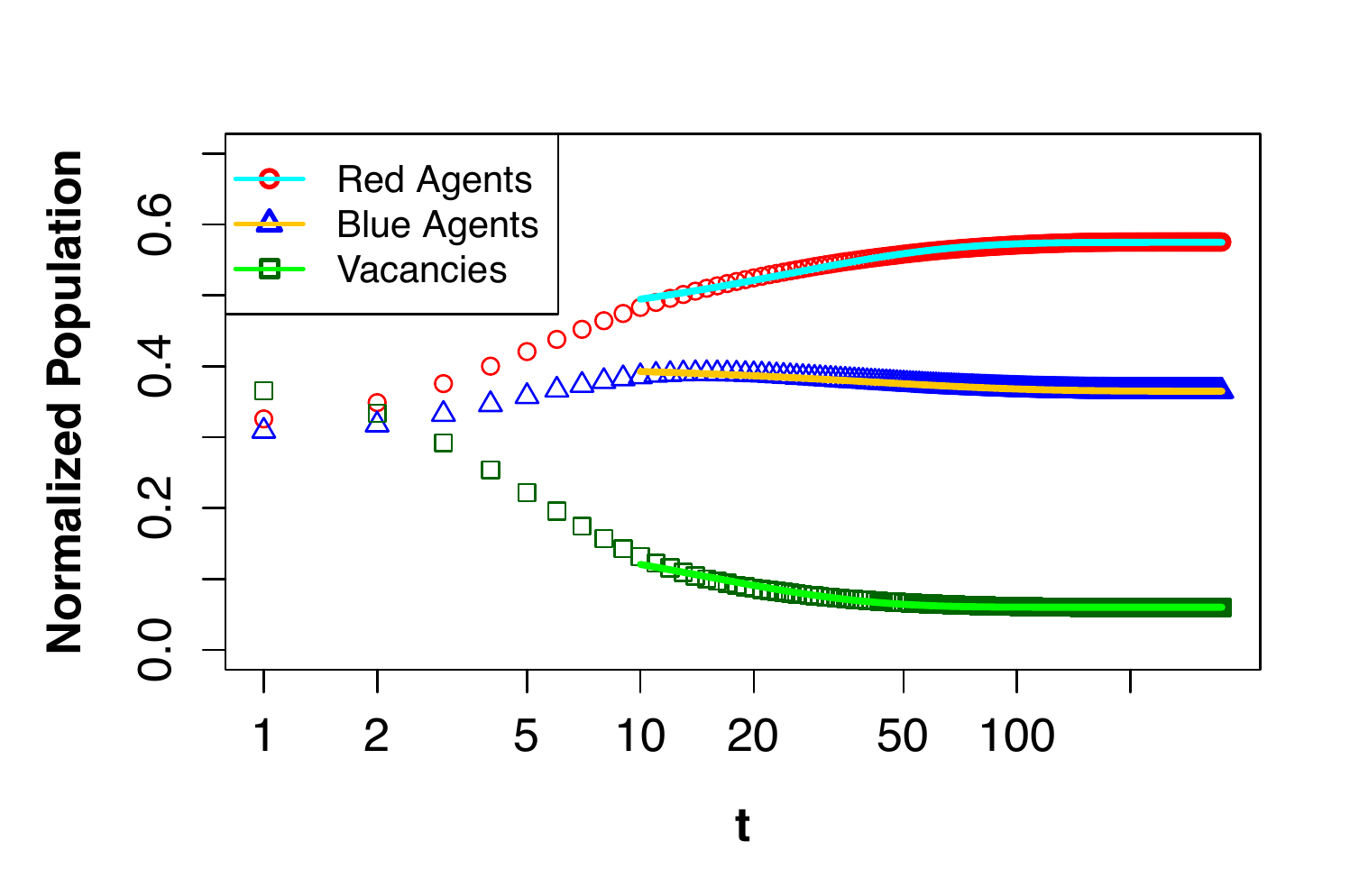}&
\includegraphics[width=5.3cm, height=3.9cm]{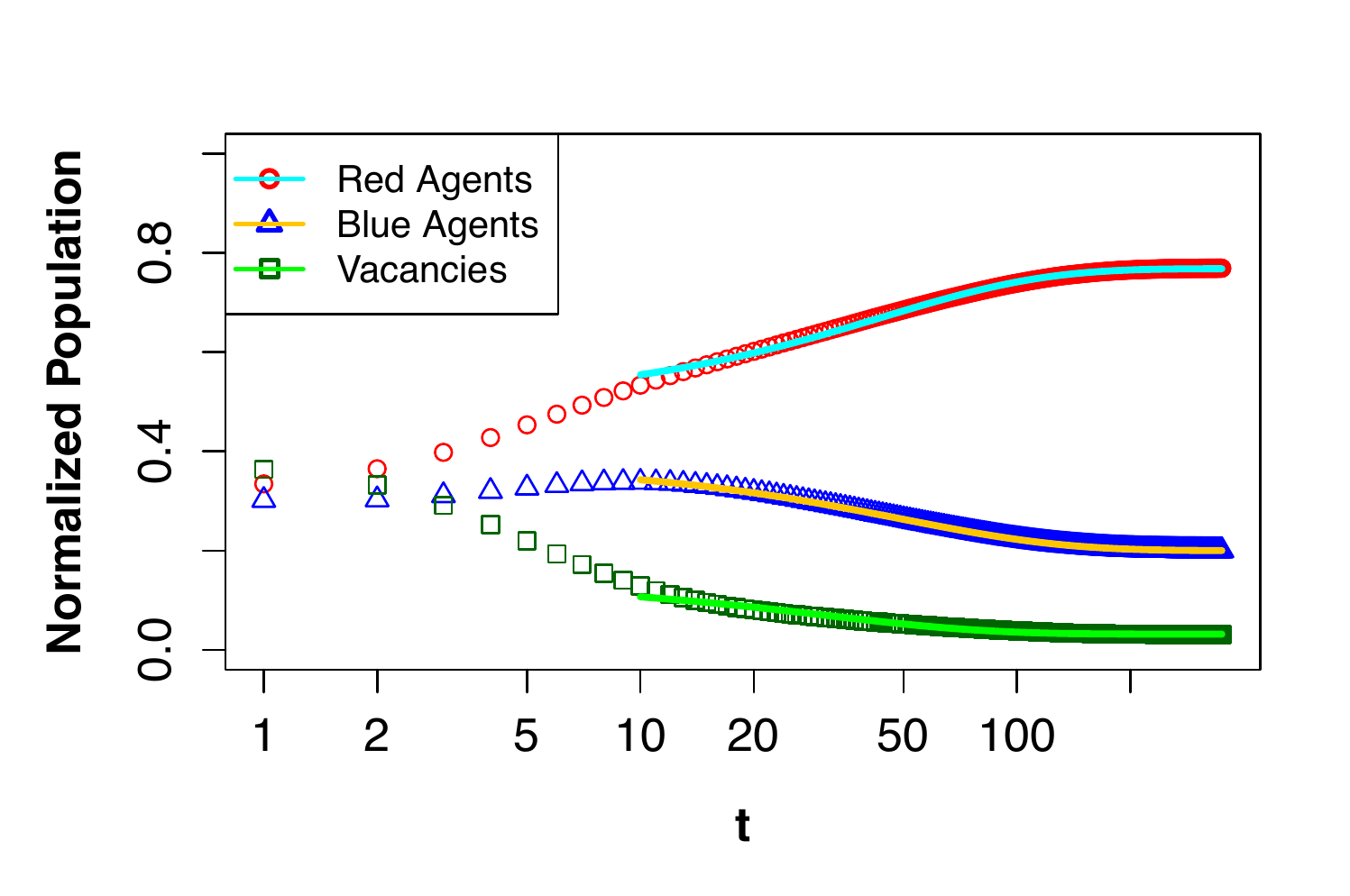}\\
a)&b)&c)\\
\end{tabular}
\caption{Evolution of normalized population during the first 350 MC steps for the suburban framework. For each graph $H$ values are $0, 0.125$ and $0.250$, from left to right. In a) cyan and orange lines overlap. Each point is averaged over 250 systems and its standard deviation calculated.  Values of $r_{adj}^2$ are over $0.97$ in all the fitted lines. The maximum standard deviation registered in figures a), b) and c) is $0.036$, $0.042$ and $0.056$, respectively. }
\label{8evol}
\end{figure}

Agent evolution can be characterized via the normalized  populations. We define $R_{p}$ as the proportion of lattice cells occupied by red agents, i.e. $N_{r}/(N \times N)$ being $N_{r}$ is the number of red agents. In other words, if $R_{p}=1$ all the lattice cells are occupied by this type of agents. Similarly, we define $R_{b}$ and  $R_{v}$ for blue agents and vacancies, respectively. These proportions can be understood as a way to normalize the lattice population as it is depicted in Fig. \ref{8evol}.

Let us take as an example the suburban framework. We start with a random configuration and equal numbers of vacancies, red and blue agents, i.e. each one of them is equal to $1/3$ of the $N \times N$ cells. The first time steps are characterized by a slow decline in the number of vacancies. This phenomenon can be observed during the first $\sim{20}$ MC steps (see Fig. \ref{8evol}). Despite the inital random state and the low tolerance value, agents start to populate the lattice. Clusters start to grow attracting neighbors of the same type, thus filling the available gaps. In their final stage, from $\sim{20}$ MC steps to the equilibrium state,  their expansion is restricted by the growth of other compact clusters and system edges. Therefore, this expansion mechanism seems to follow an exponential function: $A-B\exp \left(-t\slash t_{0}\right)$, where $A$  is the limit population, $B$ an amplitude constant and $t_{0}$ the characteristic time of the process. Contrarily, vacancies decrease as the population grows, following an inverse evolution. Finally, as $H$ increases, the economic situation results in an increase of the number of red agents while reducing the blue population as can it be seen in Fig. \ref{8evol} b) and c). Similar results can be found for other city types as it can be inferred from the data on $t_0$ values shown in Table \ref{table35}.  When there is no economic gap between red and blue agents equilibrium is reached quickly, yielding values of $t_0<12$ for all the city types. Nevertheless, when $H\neq0$ red clusters use their advantage to exert pressure over blue neighborhoods reducing their size, i.e. a gentrification process takes place. Therefore, this phenomenon increases $t_0$ values over $34$ in all cases.    

\begin{table}[H]
\begin{centering}
\begin{tabular}{|c|c|c|c|}
\hline
\ $t_0$&\textbf{$H=0$}  & \textbf{$H=0.125$} & \textbf{$H=0.250$ } \\
\hline 
\textit{Flat} & $8.6\pm 0.2$ &$70\pm 2$ & $59.0\pm 0.2$  \\
\hline 
\textit{Vertical} & $7.2\pm 0.1$ &$47.9\pm 0.6$ & $45.2\pm 0.2$  \\
\hline 
\textit{Suburban} & $7.1\pm 0.2$ &$41.0\pm 0.5$ & $48.9\pm 0.08$  \\
\hline 
\textit{Core}& $11.2\pm 0.2$ &$44.6\pm 0.8$ & $42.7\pm 0.01$  \\
\hline 
\textit{Grid} & $11.3\pm 0.2$ &$39.1\pm 0.9$ & $34.6\pm 0.1$ \\
\hline 
\end{tabular}
\par\end{centering}
\caption{Characteristic time of the process $t_0$ for blue clusters and different city types. Normalization of the grid model takes into account the presence of internal boundaries.}
\label{table35}
\end{table}

From a social perspective this suggests that ghettos are created during the initial steps of the urban formation. Under these circumstances, only the discriminated neighborhoods located in the most affordable house pricing zones are able to remain in the city during the enlargement process. This situation has occurred in American cities in which a very rapid development of the suburban areas has taken place, resulting in a simultaneous increase of residences and employment opportunities. For these reasons, the low-income housing was restricted to the urban core \cite{harvey}. Another known fact is that real cities evolve through time and adapt to its socio-economic reality. When we introduce an economic gap gentrification processes will take place, thus increasing the time needed for the city to reach a stable state.

It is also interesting to estimate the number of final blue agents as a fraction of the total population once equilibrium has been reached. These values have been compiled in Table \ref{table4}.

\begin{table}[H]
\begin{centering}
\begin{tabular}{|c|c|c|c|}
\hline
Population&\textbf{$H=0$}  & \textbf{$H=0.125$} & \textbf{$H=0.250$ } \\
\hline
\hline
\textit{Flat} & $0.47\pm0.04$ &$0.35\pm 0.06$ & $0.25\pm 0.06$  \\ 
\hline 
\textit{Vertical} & $0.47\pm0.04$ &$0.36\pm 0.04$ & $0.21\pm 0.05$  \\
\hline 
\textit{Suburban} & $0.46\pm0.04$ &$0.37\pm 0.04$ & $0.20\pm 0.05$  \\
\hline 
\textit{Core}& $0.47\pm 0.05$ &$0.34\pm 0.05$ & $0.20\pm 0.05$  \\
\hline 
\textit{Grid} & $0.47\pm 0.06$ &$0.35\pm 0.05$ & $0.23\pm 0.05$ \\
\hline 
\end{tabular}
\par\end{centering}
\caption{Normalized population of blue agents in the equilibrium. Normalization of the grid model takes into account the presence of internal boundaries.}
\label{table4}
\end{table}

Average house pricing  $\overline{D_i}$ plays a major role in values from Table \ref{table4}. As $\overline{D_i}\approx -1.14$ all the considered models exhibit an equivalent house pricing. However there is a surprising fact: boundaries do not affect greatly the results from the other models. In fact, the size of clusters decreases due to internal barriers but more ghettos are created in order to maintain a similar population. This means that the final segregated population depends on the house pricing and borders only affect the neighborhood sizes, in addition to prevent gentrification processes. This can be related to the recent results about Philadelphia (USA) in which barriers on local neighborhoods have been shown not to affect city-wide indices of segregation despite having a deep social meaning, isolating wealthy white neighborhoods and black high-poverty zones from the rest of the city \cite{barriers_2018}.

It is also possible to analyze these results from the physics perspective. This framework is equivalent to a BEG model  where the energy is being minimized, instead of the dissatisfaction (Eq. \ref{eq:4}). Obviously, $H$ maps now as a magnetic field which decreases the energy of red agents, $H(i)=-H$, and increases it for the blue ones, i.e. $H(i)=-H$. Therefore, the system tries to replace blue agents by red ones when $H>0$ yielding smaller blue clusters as $H$ grows.


\section{Conclusion}
\label{sec:conclusion}

The extended Schelling model proposed in this study allows us to deepen the understanding of segregated neighborhoods. This model is linked to the spin-1 Blume-Emery-Griffiths model in Section \ref{sec:model}) considering an open city framework where agents can leave or enter the lattice.  Besides the classical segregation terms we also consider economic features:  a financial gap between dissimilar agents $H$ and a geographical distribution of real state prices $D_i$. $D_i$ is now promoted to a field, $D(i_x,i_y)$, that depends on the location within the lattice thus defining an economic structure that we have particularized into five city frameworks: a homogeneous case,  a vertical model, which is related to coastal towns, and radial ones considering an expensive suburban zone, an expensive center and cities with an expensive center and internal boundaries. These models are defined as flat, vertical, suburban, core and grid in the same previous order. We characterized ghetto sizes, locations and  population evolution for these city types.

Independently of the considered city type, the sizes of segregated neighborhoods tend to follow a power law. The exponents are shown in Table \ref{table0} and their values lie in the range $[-1.135,-0.69]$. Therefore most ghettos are small neighborhoods while larger ones, which comprise most of the segregated population, are scarce. However, ghetto sizes show an important dependence on the city structure defined by $D_i$. All the considered city types have the same average house pricing, $\overline{D_i}\approx-1.14$, being equally attractive to settle. 

Blues cluster dimensions are related to upper cut-off $s_{max}$ parameter, interpreted as the maximum ghetto size. Nonetheless, the core and grid models exhibit smaller ghettos than the other ones, as we can see in Table \ref{table1}. This fact is specially important in the case of the grid model, due to the division of the lattice into nine isolated blocks which restrain their growth. As it was expected, when the economical gap $H$ increases, it is difficult for the segregated population i.e. the blue agents, to remain on the city. Hence, ghetto sizes decline, as Table \ref{table1} shows. 

Our model also allows us to determine the center of mass of the blue clusters and its relation with the cluster size, finding optimal ghetto locations where large segregated neighborhoods are mainly formed. Our results are expressed considering the geometry of the city type as it has been detailed in Subsection \ref{subsec:PSC}. For the flat city model, the optimal ghetto location is the center of the lattice independently of the value of $H$. In contrast, for the vertical framework,  as $H$ increases ghettos tend to be created near the system edges (Fig. \ref{4hist}) while the optimal zone moves further from the top of the town which is the most expensive area. The social interpretation is straightforward, blue agents become more and more economically handicapped and must be relocated in cheaper locations as it can be deduced from the data collected in Table  \ref{table2}. Similar results can be found for the suburban and core model. In the suburban model the growing red clusters displace the segregated neighborhoods from the central zone when the economical gap takes place. This shift is larger for the core model as it can be seen in Table  \ref{table2}. In this case, in addition to the displacement due to the red cluster expansion, the central zone is unaffordable for the segregated population. This framework, which resembles central cities around the world, explains why segregated populations are not usually located in the city center, which is usually occupied by the central business district. 

Specially interesting is the grid city model where optimal positions for the expansion of the blue clusters are located in the center of each block (see scheme from Fig. \ref{7square} a). Nevertheless, $D_i$ imposes cheaper prices as we move away from the city center and the interplay between the borders and $D_i$ yields three optimal positions for the location of ghettos (Fig. \ref{7square} b). Results show that when internal borders are considered, the effect of $H$ becomes almost negligible, and the boundaries alone can explain the structure of the segregated areas. 

The last result presented is devoted to the population evolution. All the considered frameworks exhibit a similar behavior: the system starts with a random configuration and equal number of vacancies, red and blue agents ($1/3$ from the total for each type). After a few steps, clusters of agents start to grow quickly, attracting neighbors of the same type.  However, their expansion is restricted by the growth of other compact clusters, system edges and internal boundaries, if present. Hence this saturation can be fitted by an exponential function (see Fig. \ref{8evol}) and analyzed by the characteristic time of the process $t_0$. As we can see in Table \ref{table35} the economic gap induces a drastical increase in $t_0$, due to the presence of gentrification processes in the system.  In other words, it takes longer for the city to adapt to its socio-economic reality once the segregated population is economically handicapped.

Finally, we estimated the ratio populations in the equilibrium state. These values are collected in Table \ref{table4}. We found that, despite the effect of $H$ which tends to reduce the segregated population as it increases, values from the different city models are close. To put it in another way, even in the grid model where the cluster size is limited, the final normalized segregated population is unaffected. Therefore, the final segregated population depends largely on $\overline{D_i}$ and $H$ and not on the city type, implying that ghettos are able to adapt to different city frameworks.

Further analysis should focus on variants in which agents consider city structures obtained from geographic information system (GIS). In other words, data from real towns where segregation is known to occur such as Chicago, Detroit, Rio de Janeiro, etc. Furthermore, it could be also interesting to study the distribution of the inter-event times between different economic situations (pre-pandemic and post-pandemic) and see if this theoretical framework is well adjusted to real life situations. As a conclusion, we consider that the understanding of the fundamental principles behind segregational processes and how citizens who reside in a city adapt to economic variations and different house pricing frameworks are of the utmost importance in a world where pro-segregational attitudes are commonplace.

\section*{Acknowledgments}
 We acknowledge financial support from the Spanish Government through grants PGC2018-094763-B-I00 and PID2019-105182GB-I00.

\bibliography{inho3}

\end{document}